\documentclass[manuscript]{acmart}

\usepackage{tabulary}
\usepackage{pbox}
\usepackage{gensymb} 
\AtBeginDocument{%
  \providecommand\BibTeX{{%
    \normalfont B\kern-0.5em{\scshape i\kern-0.25em b}\kern-0.8em\TeX}}}

\usepackage{array}
\newcolumntype{L}{>{\centering\arraybackslash}m{3cm}}
\usepackage{longtable}

\begin{document}

\title{A Survey on Conflict Detection in IoT-based Smart Homes}

\author{Bing Huang}
\email{bing.huang@ntu.edu.sg}
\affiliation{%
  \institution{Nanyang Technological University}
  \city{Singapore}
  \country{Singapore}
}

\author{Dipankar Chaki}
\email{dipankar.chaki@sydney.edu.au}
\affiliation{%
  \institution{The University of Sydney}
  \city{Sydney}
  \country{Australia}
}

\author{Athman Bouguettaya}
\email{athman.bouguettaya@sydney.edu.au}
\affiliation{%
  \institution{The University of Sydney}
  \city{Sydney}
  \country{Australia}
}

\author{Kwok-Yan Lam}
\email{kwokyan.lam@ntu.edu.sg}
\affiliation{%
  \institution{Nanyang Technological University}
  \city{Singapore}
  \country{Singapore}
}


\begin{abstract}

As the adoption of IoT-based smart homes continues to grow, the importance of addressing potential conflicts becomes increasingly vital for ensuring seamless functionality and user satisfaction. In this survey, we introduce a novel conflict taxonomy, complete with formal definitions of each conflict type that may arise within the smart home environment. We design an advanced conflict model to effectively categorize these conflicts, setting the stage for our in-depth review of recent research in the field. By employing our proposed model, we systematically classify conflicts and present a comprehensive overview of cutting-edge conflict detection approaches. This extensive analysis allows us to highlight similarities, clarify significant differences, and uncover prevailing trends in conflict detection techniques. In conclusion, we shed light on open issues and suggest promising avenues for future research to foster accelerated development and deployment of IoT-based smart homes, ultimately enhancing their overall performance and user experience.

\end{abstract}



\maketitle

\section{Introduction }
Recent advances in the Internet of Things (IoT) enable a growing number of physical objects (a.k.a. things) to be connected to the Internet at an unprecedented rate \cite{laghari2021review}. The rapid development of IoT is underpinned by various sensors and actuators equipped with IoT technologies \cite{atzori2010internet}. The emerging IoT technologies enable more IoT devices to be connected to the Internet.  According to a Cisco report, approximately 500 billion sensor-embedded devices will be connected to the internet by 2030 \cite{aman2020survey}. The underlying technologies, such as Near Field Communication (NFC), wired sensor networks, wireless sensor networks, and Radio Frequency Identification (RFID) tags, provide these ``things" with augmented capabilities such as networking, computing, actuating, and sensing \cite{marwedel2021embedded}. Thus, our daily life things are augmented with these capabilities and become smart. For example, everyday devices, such as Air-conditioner (AC) units, Television (TV) sets, lights, refrigerators, vacuum cleaners, and cars, are connected to the Internet in these days. IoT aims to enable connected things to \emph{see}, \emph{hear}, and \emph{sense} the physical world in the same ways that humans observe the real world. Additionally, IoT would empower these things to \emph{learn}, \emph{think}, and \emph{act} by themselves (i.e., \emph{enabling smartness}) \cite{wu2014cognitive}.

IoT technologies have been the driving force behind emerging applications, such as \emph{smart offices} \cite{mariappan2020smart}, \emph{smart campuses} \cite{sutjarittham2019experiences}, \emph{smart cities} \cite{cirillo2020smart}, \emph{smart grids} \cite{al2019iot}, and \emph{intelligent transport systems} \cite{bhardwaj2019designing}. \emph{Smart home} is another cutting-edge application of IoT. A smart home refers to any regular home that is equipped with a variety of IoT devices \cite{rashidi2013method}. These IoT devices are attached to  household objects to monitor their usage patterns. For example, a sensor (i.e., an IoT device) attached to a cup may monitor a resident's tea cup usage patterns. The IoT paradigm brings enormous opportunities to smart homes to make residents' home lives more \emph{convenient}, \emph{pleasant}, \emph{enjoyable}, \emph{entertaining}, \emph{relaxing}, and \emph{supportive} \cite{meyer2003survey}. Automating and remotely controlling IoT devices is a promising direction to fully realize the potential of smart homes.

\begin{figure*}[htbp]
\centering
\includegraphics[width=  0.9 \textwidth, height=  0.5 \textwidth]{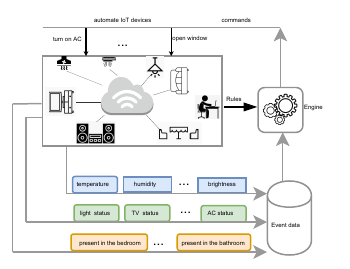}
\caption{ Smart home system}
\label{fig:smarthomesystem}
\end{figure*}

Various tools and platforms have been developed for automating lighting, heating, ventilation, and surveillance devices \cite{ur2014practical, desolda2017empowering, fogli2017imathome}. As shown in Fig. \ref{fig:smarthomesystem}, our daily life things are connected to the Internet and can be accessible by smartphones or personal computers. Residents can remotely control these devices. Composing IoT devices is a key enabler to achieving more advanced automation. The composite IoT devices are referred to as an \emph{IoT rule}. An example of such a rule is, ``If the TV is turned on, turn off the light". Once the trigger condition (i.e., the status of  IoT devices, environment properties, and resident movement) is satisfied, a corresponding command is sent to exert an action on the IoT device.  Many commercial IoT platforms, such as Samsung’s
SmartThings \cite{Samsung},   Google
Home \cite{Google}, openHAB \cite{openHAB}, Tasker \cite{Tasker}, and IFTTT \cite{IFTTT},   can support home automation.\\

Multiple IoT rules may coexist in a shared home environment and be operated simultaneously.  Concurrent execution of multiple IoT rules may cause unexpected consequences, even when each IoT rule is independently correct. This phenomenon is referred to as conflicts. \emph{Conflict} can be defined as an undesirable situation that occurs due to the invocation of a single IoT device or multiple IoT devices at the same time and location \cite{chaki2020conflict, huang2021conflict}. For example, a heating management application may request to close the window when the air-conditioning unit is turned on. At the same time, a ventilation management application may request the window to be opened to improve the room's air quality. A conflict over the window arises between these two IoT-based applications because the window cannot stay open and close at the same time and location. Conflicts usually cause undesirable consequences ranging from minor resource competence (e.g., the window is requested to be opened and closed simultaneously) to even disastrous situations (e.g., the sprinkler is disabled during the fire) \cite{halloperational}. Such conflicts may create a less comfortable home environment (e.g., an air-conditioner tries to cool the room while the window is open) if not handled properly \cite{hua2022copi}. In addition, safety issues (e.g., the front door is opened by an unauthorized entry) may occur due to conflicting situations.

There are a variety of studies on conflicts described by different terms in the literature with different contexts and assumptions, targeting different application domains. For example, conflicts refer to situations resulting from \emph{direct} or \emph{indirect} interference between IoT services \cite{huang2021conflict, chaki2020conflict}. In the literature, the term ``conflict'' in the context of IoT-based smart homes has been associated with various types of interactions and interferences. Common conflict types found include \emph{implicit interference}, where two or more rules simultaneously target multiple actuators that have contradictory effects on a shared environment property \cite{xiao2019a3id}. Another type is the \emph{feature interaction}, indicating that two services simultaneously invoke a common appliance in an incompatible manner \cite{nakamura2005feature}. Other identified conflict types encompass \emph{inter-rule vulnerabilities} \cite{wang2019charting}, \emph{interference threats} \cite{chi2020cross}, \emph{IoT app interaction threats} \cite{alhanahnah2020scalable}, \emph{inter-app interaction chain} \cite{ding2018safety}, \emph{rule interactions} \cite{yu2021tapinspector}, \emph{physical interactions} of IoT devices \cite{ding2021iotsafe}, \emph{multi-app coordination} \cite{stevens2020comparing}, \emph{attack chain} between trigger-action rules \cite{hsu2019safechain}, and conflicts arising from \emph{distinct user preferences} toward the same device \cite{oguego2021using}. A body of works have addressed various aspects of conflicts in smart homes \cite{huang2021conflict, xiao2019a3id, chaki2020conflict, goynugur2017tractable, carreira2014towards, nakamura2005feature, igaki2010modeling, nakamura2013considering}. Different approaches have been introduced to accurately and efficiently detect conflicts, such as \emph{knowledge graph-based} \cite{huang2021conflict, xiao2019a3id}, \emph{ontology-based} \cite{chaki2020conflict, goynugur2017tractable, carreira2014towards}, and \emph{object-oriented} approaches \cite{nakamura2005feature, igaki2010modeling, nakamura2013considering}.

Although conflict detection is widespread in various contexts and many related studies have been conducted, there are many inconsistencies in the area of conflict detection caused by the diverse descriptions, settings, assumptions, and application domains \cite{ma2017cityguard, preum2017preclude, tang2019conflict}. Few unified frameworks, such as IoTWatch \cite{babun2019real}, ContexIoT \cite{jia2017contexlot}, and ProvThings \cite{wang2018fear}, categorize them well, and there are no unified problem statements for conflict detection. More importantly, only a few discussions are available on the unique characteristics of conflict detection, including their problem, data, detection approaches, and conflict types \cite{chaki2020conflict, chaki2020fine, huang2021conflict}. These gaps have limited the theoretical development and practical applications of conflict detection and thus hinder the progress of intelligent home automation. Therefore,  there is a need for a systematic categorization of conflict types and representative approaches for conflict detection. To the best of our knowledge, there are three survey publications on conflict analysis in IoT systems \cite{ibrhim2021conflicts, pradeep2022conflict, resendes2014conflict}. The first survey reviewed recent work on IoT-based conflict classification by analyzing the relationships using Satisfiability Modulo Theories (SMT) and formal notations \cite{ibrhim2021conflicts}. This article  discusses IoT-based app interaction conflicts in two different scenarios:  single-user IoT app interaction conflict  and multiple-user IoT app interaction conflict. A taxonomy that defines 11 types of conflicts is used to classify IoT-based app interaction conflicts. The second article conducts a comprehensive survey of the recent work that addresses conflicts in the collective operational behavior of multiple IoT subsystems. An IoT system is characterized by a set of operational rules and safety properties. A conflict may happen when multiple IoT subsystems interact with each other and concurrently access the shared devices. In this article, three types of conflicts are discussed. The first is static conflict, which is identified through static analysis of operational rules. The second is dynamic conflict, which is detected during run-time when it happens. The third is proactive conflict, which is forecasted for a brief period in the future based on the current context at run-time. The third article explores the challenges of home and building automation systems (HBAS) when multiple users interact, leading to potential conflicts. These systems rely heavily on context detection, gathering data from sensors to infer user scenarios. The paper delves into the multi-dimensional nature of conflicts, emphasizing the need for automatic resolution or user notification.

The current surveys discuss conflicts in the contexts of IoT-based app interaction and IoT subsystems but  do not distinguish conflicts arising in a variety of \emph{contexts} in which conflicts are described  by diverse terms. More importantly, they do not systematically categorize conflict detection approaches that address conflicts in different contexts. To fill the research gaps, we present a   systematic survey on conflict classification and detection. In our survey, we consider a wider range of conflicts that arise in diverse \emph{contexts}, encompassing \emph{trigger-action rules}, \emph{IoT apps}, \emph{policies}, \emph{IoT devices}, \emph{IoT services}, and \emph{home appliances}. We systemically analyze and compare different conflict detection approaches and roughly categorize them into multiple groups. We also identify data sources used in each reviewed paper to facilitate stakeholders to discover new types of conflicts in the future. The key contributions of our work are outlined as follows:

\begin{itemize}

\item[$\bullet$] We propose a new IoT rule model based on 
which a novel  conflict taxonomy is designed. We then define different types of conflicts formally.  This conflict taxonomy is capable of capturing the unique characteristic of a conflict type  discussed in different contexts. It provides an overview of the type of conflicts concerned in different studies, aiming at establishing a common understanding of the same conflict type defined in different terms.

\item[$\bullet$] We review recent work extensively and systematically classify conflicts in IoT-based environments using the   conflict taxonomy. In detail, we identify conflicts arising in the context of trigger-action rules, IoT apps, policies, IoT devices, IoT services, and home appliances and map them onto our taxonomy. The purpose of the systemic classification is to provide an overview of what types of conflicts have been addressed in different contexts. 

\item[$\bullet$] We provide a comprehensive literature review of state-of-the-art conflict detection approaches. In particular, we categorize these approaches into five main groups, including graph-based, object-oriented, formal rule modeling, model checking, and other approaches. A systematic classification and comparison of conflict detection approaches are made to provide an overall view of how conflicts have been detected and what progress has been achieved.

\item[$\bullet$] We discuss some open issues in the field of conflict detection and point out potential future research.

\end{itemize}

The organization of the rest of the paper is as follows. Section 2 provides the background on conflict regarding the IoT-based smart home environment, smart home automation,  and  conflict definitions in  IoT-based smart home scenarios. In section 3,   a novel conflict taxonomy is presented with a formalization  of each type of conflict based on a new IoT rule model. Our conflict taxonomy serves as a bridge that establishes a common understanding of conflicts addressed in diverse IoT contexts. As a result, it is used as a tool to categorize a variety of conflicts discussed in recent literature, which is presented in Section 4.  In section 5, a systematic classification and comparison of conflict detection approaches are made to provide an overview on how conflicts are detected in different contexts.  Each class of approaches has been briefly introduced with key technical details to provide an in-depth understanding of the progress achieved for conflict detection.  Section 6 provides some insights and future research directions. We outline the threats to validity in section 7. Finally, section 8 concludes the paper.

\section{Setting the Stage: Conflicts and IoT-based Smart Homes}
We provide the notion of the \emph{IoT-based smart home environment} (section 2.1) and \emph{smart home automation} (section 2.2) to better understand IoT conflicts in smart homes. We then provide the latest \emph{conflict definitions} in different works to demystify the concept of conflict in section 2.3.

\subsection{IoT-based Smart Home Environment}
The Internet has transferred from traditional linking computers to people's current interconnection and the emerging connection of things over the decades \cite{sheng2017managing}. The Internet of Things (IoT) aims to connect billions of things and augment them with capabilities of seeing, hearing, sensing, and responding to the real world. Our everyday things, such as shoes, clothes, TV, and lights, are becoming part of the Internet and more intelligent \cite{asghari2019internet}. Numerous smart devices, including infrared motion sensors, switch sensors, pressure sensors, accelerators, wearable sensors,  humidity and temperature sensors, embedded systems, and actuators, drive the rapid development of the Internet of Things \cite{whitmore2015internet}. In addition, the emerging IoT relies on a multitude of enabling technologies, including wireless sensor networks (WSN), identification and tracking technologies, distributed systems, and enhanced communication protocols \cite{al2015internet}.

The prevalence of the IoT brings  unprecedented opportunities in various domains, including smart cities, digital health, smart transportation, smart logistics, smart campuses, and smart homes \cite{bandyopadhyay2011internet, zeinab2017internet}. In this paper, we focus on the smart home domain. In the near future, common appliances such as TV, fridge, microwave, window, and plants may be connected to the Internet \cite{sheng2017managing}. One of the ultimate goals of smart homes is to enhance residents' living experience by creating an  intelligent home environment. To reach this goal, real-world testbeds, such as the MavHome project, CASAS project, Aware Home, iDorm, and Gator Tech Smart House, have made some attempts to realize such a smart home environment \cite{rashidi2010discovering}. In addition, many IoT-based applications have been developed, including activity recognition, health assistance, environment monitoring, and home automation \cite{ni2017securing, dahmen2017smart}. In particular, composing IoT devices into trigger-action rules is  one of the promising means of automating smart homes.  For example, the resident can customize IoT-based applications using the trigger-action paradigm \footnote{https://ifttt.com/home} to mash up multiple IoT devices. An example of such a rule is, ``If the TV is turned on, turn off the light''. Furthermore, an IoT device can communicate with another device to automate it by sending messages \cite{nakamura2008constructing}. We provide more details about smart home automation in section 2.2.

\subsection{Smart Home Automation}
Smart homes aim to enhance people's life quality \cite{meyer2003survey, lee2020critical, singh2019energy}. Personalizing the smart home environment by the End-User-Development paradigm is one of the promising solutions to achieve such goals. End-User Development (EUD) focuses on providing technologies and tools and putting the IoT-based applications/services development in the hands of residents who are most familiar with the actual needs but have limited programming skills \cite{ghiani2017personalization}. By employing EUD tools, non-technical residents can personalize their smart environment easily and autonomously. For example, a resident can easily customize his/her behavior, such as ``IF I am watching TV, THEN turn off the light and close the curtain" to create a theater-like atmosphere. So far, many approaches have been proposed to facilitate end-user development, such as the natural language approach  \cite{perera2015natural}, the model-driven approach \cite{trouilhet2021model}, the semantic approach \cite{roffarello2018end}, and the semantic conversational search and recommendation approach \cite{corno2021users}. EUD tools are also available for personalizing IoT-based applications/services such as IFTTT \cite{IFTTT}, Tasker \cite{Tasker},   TAP(Trigger-Action programming) \cite{zhao2020visualizing}, and TARE (Trigger-Action Rule Editor) \cite{ghiani2017personalization}. The simplicity and expressiveness of these tools make them the most widely used representations. A variety of commercial platforms are also available to support programmers in developing IoT-based applications that implement proper functionality on IoT devices. The open-source openHAB platform, SmartThings from Samsung, and Android Things from Google are representatives of popular commercial platforms \cite{openHAB, Samsung, Google}. In addition, a smart home system called WITSCare has been developed to help older people independently live in their own homes \cite{yao2016context}. The system identifies the IoT-enabled devices associated with contexts (i.e., residents' locations, activities, and their interactions with home appliances) and abstracts them as services. The abstracted services are exposed in the form of RESTful APIs and further represent the APIs as graphical icons. Residents can personalize their smart homes by creating complex rules in a drag-and-drop fashion without any programming efforts.

\subsection{Conflict Definitions in IoT Scenarios}
A conflict generally refers to a   disagreement between opposing attitudes, ideas, values, or needs  \cite{wang2011development}. Conflict involving an application or user is defined   as \textit{`` a context change that leads to a state of the environment which is considered inadmissible by the application or user"} \cite{tuttlies2007comity}. Conflict is widely discussed in various domains, such as smart cities \cite{ma2016detection, ma2017cityguard}, smart homes \cite{resendes2014conflict},  software systems \cite{hepner2006patterns}, autonomous cars \cite{abdessalem2018testing},  aircraft transportation \cite{jilkov2018multiple}, human-computer interaction \cite{miandashti2020empirical} and so on. This survey focuses on conflicts occurring in the domain of smart homes. With the help of EUD tools and technologies, residents can quickly and easily develop IoT-based apps/services based on their own needs in a shared smart home environment. In this regard, there will be an   increasing number of IoT-based apps/services developed by the residents. As a result, conflicts may arise when simultaneously running multiple IoT-based apps/services. To the best of our knowledge, conflict can be defined in different ways, and in general, it indicates some  undesirable situations. In this regard, conflict is referred to as different terminologies but has a similar meaning in the papers we have surveyed. For example, conflicts refer to  \emph{interference} between IoT services \cite{huang2021conflict, chaki2020conflict} and  \emph{implicit interference} between trigger-action rules \cite{xiao2019a3id}. It is also termed as  feature interaction between appliances \cite{nakamura2005feature}. In the IoT security area, some research works define conflict as  \emph{inter-rule vulnerabilities}   \cite{wang2019charting},  \emph{interference threats}  \cite{chi2020cross}, \emph{IoT app interaction threats} \cite{alhanahnah2020scalable},  \emph{inter-app interaction chain}   \cite{ding2018safety},   \emph{rule interactions} \cite{yu2021tapinspector},  \emph{physical interactions}     \cite{ding2021iotsafe},   \emph{multi-app coordination}   \cite{stevens2020comparing}, and   \emph{attack chain} between trigger-action rules \cite{hsu2019safechain}. Therefore, it is necessary to clarify these terminologies and establish a common understanding of conflicts that have similar meanings but are defined in different terms.

\subsection{Unique Characteristics of Conflicts in Smart Homes}
The unique characteristics of conflicts in smart homes stem from their operational scale, user interactions, and the balance between comfort and efficiency. Smart homes operate on a smaller scale, leading to conflicts that arise from the intimate interplay of a limited number of devices tailored to individual or family needs \cite{garg2022social}. The diverse user interactions in smart homes, characterized by varying preferences and tech-savviness of household members, further contribute to potential conflicts \cite{brich2017exploring}. Additionally, as devices in smart homes are often customized for individual preferences, conflicts can emerge when trying to strike a balance between ensuring user comfort and maintaining operational efficiency \cite{strengers2020pursuing}.

Furthermore, the nature of safety concerns in smart homes, which prioritize physical safety, security, and user comfort, can lead to unique safety-driven conflicts \cite{mocrii2018iot}. The varied and often unpredictable contexts in which smart home devices operate introduce another layer of complexity, resulting in contextual conflicts as devices might receive conflicting signals or data from their environments \cite{perera2013context}. Emphasizing personal privacy in smart homes can also lead to conflicts, especially when devices have overlapping functionalities \cite{islam2012security}. Lastly, the emphasis on security in smart homes can lead to conflicts, especially when unauthorized access is detected or when devices operate with conflicting security protocols \cite{sikder2020kratos}. These characteristics underscore the intricate nature of conflicts in smart homes, emphasizing the need for sophisticated conflict detection and resolution mechanisms tailored to the unique challenges of residential environments.

\section{Conflict Taxonomy}
We first introduce the notion of the IoT device, which is defined by a variety of properties and methods. Then, we define the IoT rule, which aims to automate IoT devices using the trigger-action  paradigm. Lastly, we present a conflict taxonomy and formalize different types of conflicts  based on the IoT rule model.

\subsection{IoT Rule Model}  
We adopt the object-oriented approach   to model an IoT device due to its simplicity   in capturing different levels of properties \cite{nakamura2013considering}. The IoT device is described by a set of methods and properties.  We introduce a new IoT rule model to describe  how IoT devices can be combined to fulfill automation. 

\textbf{\noindent IoT device}. An IoT device is  described by a tuple $D = < id, \alpha, \beta, Eff,  ST, \delta  > $ where:
\begin{itemize}
\item[$\bullet$] $id$ is a unique identifier for the smart object. 
\item[$\bullet$] $\alpha = <\alpha_s, \alpha_e>$ are a pair of methods where $\alpha_s$ is an action for activating the object and $\alpha_e$ is an action for deactivating the object.  
\item[$\bullet$] $\beta = \{ \beta_1, \beta_2,..., \beta_n \}$ is a set of functionalities or methods offered by the object. For each $\beta_i$, it is denoted as $\beta_i = < name_i, input_i >$, where $name_i$ is the function name and $input_i$ is the input parameter of the functionality. 

\item[$\bullet$] $Eff = \{ env_1, env_2,.., env_n\}$ is a set of  environment effects associated with the device $D$. An environment property $env_i$ is  impacted if   the usage of the object $D$ changes the value of $env_i$.  

\item[$\bullet$] $ST = \{st_1, st_2,..., st_n\}$ is a set of possible states of a device.
\item[$\bullet$] $\delta$ is a   function describing the relation between device methods and states and environment effects. It is denoted as $\delta: st_i \xrightarrow {f\in \alpha \cup  \beta} st_j \odot env_k$ meaning that executing the method $f$ would change the device state from $st_i$   to $st_j$ and change the environment property $env_k$. 
 
\end{itemize}

For example, a fan  can be represented as $\langle$  $\alpha = \langle fan.on(),  fan.off()\rangle$, $\beta = \{fan.setFanSpeed(speed)\}$, $Eff = \{temperature\}$, $ST = (ON, OFF)$, $\delta: OFF \xrightarrow {fan.on()} ON \odot \{temperature\}$ $\rangle$. A window can be represented as $\langle$  $\alpha = \langle window.open(),  window.close()\rangle$, $\beta = \{\}$, $Eff = \{temperature, humidity\}$, $ST = (OPEN, CLOSE)$, $\delta: CLOSE \xrightarrow {window.open()} OPEN \odot \{temperature, humidity\}$ $\rangle$. 

\textbf{\noindent IoT rule}. An IoT rule $r$ is represented by $trig\rightarrow f$ where:
\begin{itemize}
\item[$\bullet$] $trig = trig_1 \wedge trig_2 \wedge...trig_k$ is referred to as triggers which is a set of conditions  connected by the logic operators \textbf{AND}. The trigger condition $trig_i$ is one of the possible IoT device states $ST$, environment properties $Env$ (i.e., we consider seven common environment properties, these are $Env $= \{temperature,  brightness, humidity, smoke, sound, humidity,  and  CO2\}), time $T$, and resident's presence in a location $Loc$,  that is $trig_i \subseteq (ST \cup Env \cup T \cup Loc )$.   
\item[$\bullet$] $f$ is referred to as an action  that is one  of IoT device functionalities to be executed with $f \in (\alpha \cup \beta)$.    
\end{itemize}
For example, the IoT rule `` fan.state = ON  $\rightarrow$  window.open()''  means opening the window when the fan is active. According to the window example mentioned above, opening the window may change room temperature and humidity.

\subsection{Conflict Taxonomy}
\begin{figure*}[t]
\centering
\includegraphics[width=\textwidth]{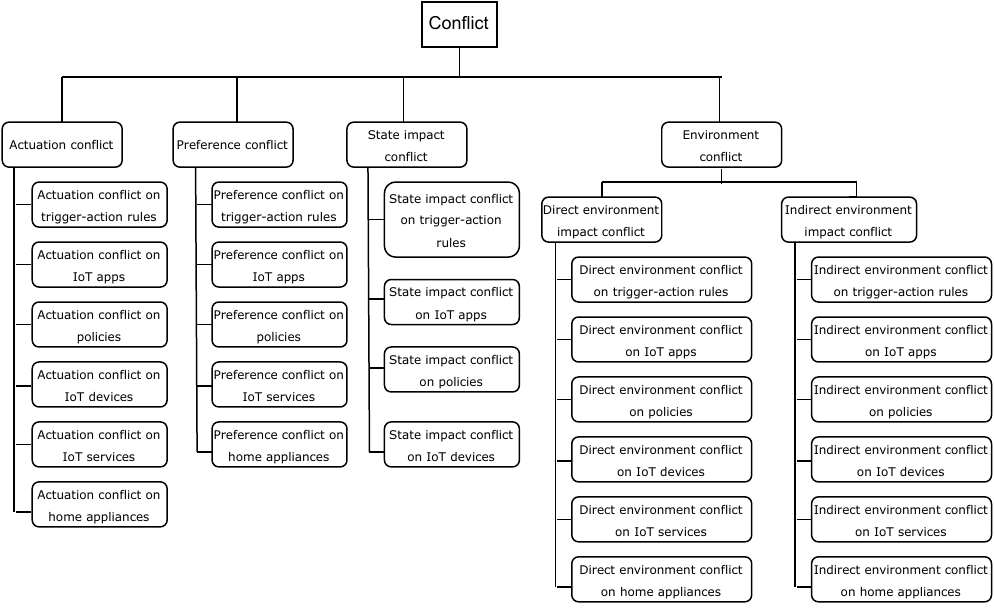}
\caption{ Conflict taxonomy}
\label{fig:taxonomy}
\end{figure*}

This section provides a conflict taxonomy and formally defines different types of conflicts based on the IoT rule model. As shown in Fig. \ref{fig:taxonomy}, the conflict taxonomy has four main categories of conflicts. They are (i) Actuation Conflict, (ii) Preference Conflict, (iii) State Impact Conflict, and (iv)  Environment   Conflict. The Environment Conflict category has two sub-types of conflicts including Direct Environment Impact Conflict and Indirect Environment Impact Conflict. In total, we have five types of conflicts under the four categories.  For each type of conflict, it is discussed in a variety of contexts such as trigger-action rules, IoT apps, policies, IoT devices, IoT services, and home appliances. Therefore, our taxonomy further develops sub-types of conflicts based on these contexts. We provide a formal definition of different types of conflicts and use examples to illustrate these conflicts as shown in Fig. \ref{fig:examples}. We formalize  the conflicts based on our IoT rule model as follows:

\begin{figure*}[b]
\centering
\includegraphics[width= \textwidth]{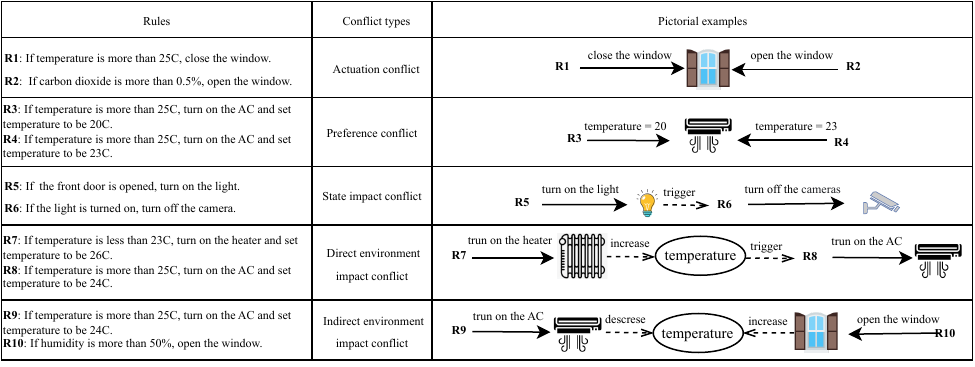}
\caption{  Illustration of different types of conflict  }
\label{fig:examples}
\end{figure*}

\noindent \textbf{\noindent Actuation Conflict:} An actuation conflict refers to a situation that two IoT rules interact with a shared device in  an incompatible way at the same time. For example, an IoT rule requests to turn  on the air-conditioning for cooling the living room while another rule requests to turn off the same air-conditioning. An actuation conflict happens in this situation as the air-conditioning cannot be turned on and off at the same time. Formally, given two IoT rules $trig^i\rightarrow f^i$ and  $trig^j\rightarrow f^j$, a possible actuation conflict may arise if one rule aims to activate a device $D$ while another rule aims to deactivate the same device. The following condition is satisfied:

\begin{equation}
 ( f^i =  \alpha_s^D )\land ( f^j = \alpha_e^D )   
\end{equation}

\noindent \textbf{\noindent Preference Conflict:} A Preference Conflict refers to the situation that two IoT rules interact  with a shared device simultaneously and  access the same method with   different input parameters. For example, a rule turns on the air-conditioning and sets the temperature to 20 degrees to cool the living room. Another rule requests to set the temperature to 25 degrees. As a result, a conflict happens as the temperature of the air-conditioning unit cannot be set to 20 and 25 degrees at the same time. Formally, given two IoT rules $trig^i\rightarrow f^i$ and  $trig^j\rightarrow f^j$, a Preference Conflict may arise if the $f^i$  and $f^j$ are the same method with different inputs. The following conditions are satisfied:
 
\begin{equation}
  (f^i = f^j) \wedge (input_i \neq input_j) 
\end{equation}

\noindent \textbf{\noindent State Impact Conflict:} A State Impact Conflict arises if a particular state of a device invokes or disables another IoT rule to execute accidentally. For example, the rule  ``If the front door is opened, turn on the light'' may trigger another rule ``If the light is turned on, turn off the camera''. Formally,  given two IoT rules $trig^i\rightarrow f^i$ and  $trig^j\rightarrow f^j$, the former rule transforms the device $D_i$ into a state $st_i$ by executing the method $f^i$. The device state $st_i$ is one of the triggering conditions of the latter rule. The following conditions should hold.

\begin{equation}
 st_i \in trig_j
\end{equation}

\noindent \textbf{\noindent Direct Environment Impact Conflict:} A Direct Environment Impact Conflict  happens if the environmental impact of using  a device triggers or disables the execution of another rule accidentally. For example, a rule turns on the humidifier to  make  the air   more humid in the living room. The air humidity increases after turning on the humidifier for a while. As a result, the increasing air humidity may  activate the dehumidifier  automatically to dehumidify the air. Formally, given two IoT rules $trig^i\rightarrow f^i$ and  $trig^j\rightarrow f^j$, the involved device $D_i$ in the former rule impacts a set of environment properties $Eff_i = \{ env_1, env_2,.., env_n\}$ after invoking the method $f^i$. These environmental properties may be the triggering conditions of the latter rule. The following condition is satisfied.

\begin{equation}
Eff_i \cap  trig_j \neq  \phi
\end{equation}

\noindent \textbf{\noindent Indirect Environment Impact Conflict:} An Indirect Environment Impact Conflict refers to the situation that two rules interact with two  different devices while the two devices impact the shared environment property. For example, a rule requests to use the air-conditioning to cool the living room in hot summer while another rule opens the window to let outside fresh air come in. A conflict may occur because opening the window would let hot air come in and   impact the air-conditioning's cooling effect.  Formally, given two IoT rules $trig^i\rightarrow f^i$ and  $trig^j\rightarrow f^j$, they access the device $D_i$ and $D_j$. The environment properties after executing $f^i $ and $f^j$ are $Eff_i = \{ env_1, env_2,.., env_n\}$ and $Eff_j = \{ env_1, env_2,.., env_m\}$, respectively. The following conditions should hold:

\begin{equation}
   (D_i\neq D_j)  \wedge (Eff_i \cap   Eff_j \neq  \phi)
\end{equation} 

\section{Conflict Classification}
In this section, we classify conflicts discussed in the existing works using our conflict taxonomy as follows. The classification result is summarized in Table \ref{mapping}. We analyze the conflicts in the existing works and generalize them into our conflict taxonomy based on the formal definitions. We also  collect the associated example used in the surveyed papers to elicit the meaning of each type of conflict. For a better understanding, we have developed a comparison protocol as follows:

\begin{itemize}

    \item Develop the conflict taxonomy: We created a comprehensive taxonomy of conflicts in IoT-based smart homes, including Actuation, Preference, State Impact, Direct Environment Impact, and Indirect Environment Impact conflicts. We identify these conflicts arising in different contexts including trigger-action rules, IoT apps, policies, IoT devices, IoT services, and home appliances.

    \item Identify relevant publications: We searched for and selected publications that discuss or address conflicts in IoT-based smart homes. These publications should provide information about the types of conflicts and the contexts (specific situations or scenarios) in which they occur.

    \item Extract conflict classification information: For each publication, we extracted information about the types of conflicts discussed and their corresponding classification within our proposed taxonomy, as well as the contexts in which these conflicts occur.

    \item Populate the table: We compiled the extracted information in a table format, allowing for a clear visualization of how conflicts in the related work are classified using our taxonomy and their contexts.
    
\end{itemize}

It's necessary  to understand the context in which conflicts arise. We have identified six main types of contexts in which conflicts are investigated in IoT-based systems. These contexts are trigger-action rules,  IoT apps,  policies, IoT devices, IoT services, and home appliances. Modeling these contexts is a key prerequisite to identifying conflicts. Specifically, the trigger-action rule can be described by a condition and an action (e.g., ``If temperature > 25C $\to$ airconditioner.on()''). An IoT service is usually modeled by a set of functionalities. Similarly, an IoT device or an appliance can also be modeled by a set of methods using the object-oriented approach. For example, a light can be described by a tuple ``$<$ light.on(), light.off() $>$''. An IoT app is a piece of program that implement certain functionalities. A policy is usually described using plain text (e.g., ``The door must always be locked when the user is not at home''.).

We use our taxonomy to classify various conflicts studied in the context  of  trigger-action rules \cite{ibrhim2020formal, sun2014conflict, wang2019charting, xiao2019a3id, yu2021tapinspector}, IoT apps \cite{chi2020cross, alhanahnah2020scalable, celik2019iotguard,  hamza2022hsas, li2020diac,   shen2017systematic, stevens2020comparing, trimananda2020understanding}, policies \cite{liu2019remediot, nagendra2019viscr, pradeep2021automating, goynugur2017tractable, shehata2007using},    IoT devices \cite{alharithi2019detecting, chen2019multi, ding2021iotsafe, ding2018safety}, IoT services \cite{chaki2020fine, chaki2020conflict, huang2021conflict}, and home appliances  \cite{leelaprute2008detecting, nakamura2005feature}.

\subsection{Actuation Conflict}  
Actuation Conflict is addressed in the majority of the surveyed papers, as shown in Table 1. We identify Actuation Conflict occurring in the context  of trigger-action rules \cite{ibrhim2020formal, sun2014conflict, wang2019charting,  yu2021tapinspector}, IoT apps \cite{chi2020cross, alhanahnah2020scalable, celik2019iotguard,  hamza2022hsas, li2020diac,   shen2017systematic,  trimananda2020understanding, yagita2015application}, policies \cite{liu2019remediot, nagendra2019viscr, pradeep2021automating, goynugur2017tractable},    IoT devices \cite{alharithi2019detecting}, IoT services \cite{chaki2020fine, chaki2020conflict, huang2021conflict}, and home appliances  \cite{leelaprute2008detecting, nakamura2005feature}.

\subsubsection{Actuation Conflict on Trigger-Action Rules}~\\
A typical trigger-action rule can be defined by a set of triggering conditions connected by the logic operators and associated actuators. An example of such a rule is ``temperature > 25C $\to$  airconditioner.on()'' meaning if temperature is higher than 25C, then turning on the airconditioner. Actuation Conflict may arise when multiple rules request to access the shared actuators. The majority of studies consider Actuation Conflict under different circumstances. We review the actuation conflict addressed in the context of trigger-action rules as follows. 

 A framework called Rule Verification Framework is introduced to detect and resolve conflicts between trigger-action rules \cite{ibrhim2020formal}. In particular, the \emph{Redundancy Conflict (RC)} and the \emph{Condition Inconsistency Conflict (CIC)}, as well as the \emph{Non-Specified Conflict (NSC)}, can be generalized to our Actuation Conflict because two rules act on the shared device no matter what their conditions are. Specifically, the RC refers to that two rules have the same action while one rule's condition is a subset of another rule's condition. Similarly, the CIC refers to a situation where two rules have the same action while their conditions are contradicting. Likewise, the NSC refers to the situation where two rules with distinct conditions are activated simultaneously, and their consequences on the system are contradicting, i.e., one rule tries to close the window when PM 2.5 is more than $80g/m^3$ while another rule opens the window when it is 7 am.

A variety of conflicts among rules are discussed in \cite{sun2014conflict}. The \emph{Shadow Conflict} and the \emph{Execution Conflict} describe typical scenarios of Actuation Conflict in the context of rule-based systems. In particular, the \emph{Shadow Conflict} describes the situation that two rules share the same actuator while the trigger condition of one rule is contained by the other rule. Similar to the Shadow conflict, the \emph{Execution Conflict} describes the situation that two rules take contrary actions on the same actuator while   the trigger condition of one rule is contained by the other rule. Various types of  inter-rule vulnerabilities are discussed in \cite{wang2019charting}.  The \emph{condition bypass}, \emph{action revert}, and \emph{action conflict} describe different scenarios of Actuation Conflict between two rules. The \emph{condition bypass} refers to the situation that two rules take the same actions on a shared object while the condition of a rule is the subset of another rule’s condition.  Both ``action revert'' and ``action conflict'' describes the situation that two rules take incompatible actions on a shared object at the same time.

A system called TAPInspector is developed  to detect rule interactions \cite{yu2021tapinspector}. The \emph{Actuation Duplication},  \emph{Action Conflict}, \emph{Action Overriding}, and \emph{Action Breaking} can be mapped to our Actuation Conflict category because they describe the situation where two rules act on the shared device at the same time. Specifically, an Action Duplication refers to a situation where two rules are activated simultaneously with repeated actions. For example, two rules try to turn on the light at the same time. Much the same as Action Duplication, the    Action Conflict refers to a situation where two rules are activated simultaneously with conflicting actions. For example, two rules try to turn on and turn off the same light simultaneously. Similarly, an Action Overriding refers to a situation where one rule  overrides the effect of another rule. For example, the rule ``When motion is detected, turn on the light'' has Action Conflict with another rule ``When motion is detected over one minute, turn off the light''. Thus, the Action Overriding can also be generalized to our Actuation Conflict because two rules act on the same device within a short period of time. An Action Breaking refers to a situation where  one rule breaks the action in the progress of another rule. For example, the rule ``Turn on fan for 15 minutes when CO2 >1000ppm'' has Action Breaking conflict with another rule ``At noon turn your fan on for 15 minutes''.  If the latter rule turns on the fan for a few minutes (not more than 15), the fan will be turned off by the rule and then turned on again by the former rule, which leads to a conflict. Another special case of Actuation Conflict is called   Condition Blocking. It refers to a situation where one rule blocks the condition of another rule. For example, the rule ``If humidity is below 38\%, start the watering system'' has the Condition Blocking conflict with the rule ``Over 5 minutes after humidity is below 38\% while the watering system is turned on, close the watering system''. Hence, the Condition Blocking can be generalized to our Actuation Conflict category because   two rules request incompatible actions at the same time (e.g., starting the watering system vs. stopping the watering system).

\subsubsection{Actuation Conflict on IoT Apps}~\\
An IoT app can be considered as the implementation of a trigger-action rule. It can be written in different programming languages. For example, SmartApp on the SmartThing platform can be programmed in Groovy language. Some studies focus on discovering Actuation Conflict between IoT apps via static code analysis. We review these Actuation Conflicts addressed in the IoT apps context. 

Seven types of IoT app interaction threats are addressed in \cite{alhanahnah2020scalable}. Two types of these interaction threats (i.e., \emph{action-action (repeat)}, \emph{exclusive event coordination}) capture different scenarios of Actuation Conflict in the context of multiple IoT apps. In detail, the  action-action (repeat) interaction threat refers to the situation where two apps act on the same device with the same value. Likewise, the exclusive event coordination describes the situation where two apps act on the same device even with exclusive triggering events. Additionally, two types of interactions characterize different scenarios of the Actuation Conflict, which may cause unsafe and insecure interactions between IoT apps \cite{celik2019iotguard}.  The first type  of interaction is that   multiple apps can change the same attribute of a  device at the same time.   For example,  an app turns off the alarm when the resident is detected, while another app  turns on the alarm when it detects the motion of the resident. The second type of interaction refers to a similar situation where apps could respond to the same event in conflicting ways. For example,  when a contact sensor is open, one app turns on the light, while another app turns off the light.

A variety of \emph{interference threats} are addressed in  \cite{chi2020cross}. In particular, the \emph{Action-interference} threat  refers to the situation where two apps operate on the same actuator by issuing conflicting commands such as opening/closing the window. In this regard, the action interference threats can be mapped to our Actuation Conflict category. Additionally, a framework named   HSAS-MD analyzer  is presented to detect the action conflicts between IoT apps \cite{hamza2022hsas}. The action conflict, also referred to as malware, can be mapped to our Actuation Conflict category. It refers to a scenario in which two apps that operate on the same device but have different values—for example, simultaneously turning on or off the light—conflict with each other's actions. Similarly,  the Actuation Conflict is  referred to as the \emph{Feature Conflict } in the context of app interaction \cite{trimananda2020understanding}. When two apps attempt to update the same device state with incompatible values, this is known as a \emph{feature conflict} (e.g., an app attempts to door-unlock during a fire while another app could potentially door-lock).  

The \emph{weak conflict} discussed in \cite{li2020diac, shen2017systematic} can be generalized to our Actuation Conflict category. A weak conflict occurs when numerous apps control the same actuator in distinct ways without any of the operations of the apps being disabled. For instance, an energy-saving app turns off the light when there hasn't been any movement detected in the last two minutes, whereas a security app makes the light alternately stay on and off periodically (once every half hour for each state) when the resident leaves his home to simulate the presence of people. The security app and energy-saving software are in a weak conflict because the latter cannot imitate the presence of people in the house because the lights are typically off. Close to the above mentioned weak conflict, actuator conflicts   may occur if two or more apps are running simultaneously, especially when they try to use a single actuator \cite{yagita2015application}. The actuator conflict discussed in the study can be mapped to our Actuation Conflict category. For example, the Wind Blower app opens a window when the status of the wind sensor is true. The SecureWindow app closes the window when the status of humanAct sensor is asleep or away. These two apps may request to open and close the same window concurrently, resulting in a conflict.

\subsubsection{Actuation Conflict on Policies}~\\
 Conflicts may exist among multiple policies. A policy can be considered as a rule specification  of a particular system with respect to some objective function such as safety or utility \cite{liu2019remediot}. It can be simply treated as the specification for a trigger-action rule or some safety properties described in plain texts. An example policy may aim at preventing racing conditions between two applications that may turn
a device on or off. Racing events investigated in \cite{liu2019remediot} is a type of conflict that violates certain policies. They can be mapped to our Actuation Conflict.  For example, a policy that turns on the fan when carbon dioxide density is conflicting with another policy that turns off the fan when the temperature is low. Likewise, an approach called VISCR is introduced to detect a variety of conflicts   between policies \cite{nagendra2019viscr}. The \emph{potential conflict} in this study captures the characteristics of our   Actuation Conflict.  It can be loosely defined as   two policies having conflicting actions and lacking  specific temporal attributes among these policies. For example, a policy tries to close the  window  in case of a specific outdoor temperature, while another policy opens the window in case of rain and humidity. It is impossible to close and open the window at the same time. Similarly, the \emph{direct conflict} refers to the situation where two operational policies have opposing actions for the same actuator \cite{pradeep2021automating}. An example of a direct conflict is if two rules ask for opening and closing the door at the same time. Thus, the direct  conflict can be mapped to our Actuation Conflict type. Likewise, policy conflict is also addressed in \cite{goynugur2017tractable}. Two policies have conflicts if they concurrently oblige/prohibit the same action on the same device, which is modelled by our Actuation Conflict. For example, one policy states that ``notify when the doorbell is pressed by an audio alarm, and if
not responded in 15 minutes, send a message to the registered mobile phone.''. Another policy claims that `` no device should make noise when the baby is sleeping''. The two policies may have conflicting actions on the doorbell.

\subsubsection{Actuation Conflict on IoT Devices}~\\
Actuation Conflicts are discussed in the context of IoT devices in smart homes. In  \cite{alharithi2019detecting},  both \emph{direct} and \emph{indirect} conflicts are discussed and defined conceptually. Direct conflicts are further divided into three subcategories, including \emph{overwrite conflict}, \emph{opposite conflict}, and \emph{environmental conflict}.  The three subcategories of conflicts can be mapped to our Actuation Conflict because they characterize the situation where   incompatible actions are exerted on a shared device at the same time.  Specifically,  the opposite conflict refers to the situation that two rules exert two contrary actions on a shared device,  e.g., turning on the light and turning off the light by two respective rules. Likewise, the overwrite conflict refers to the situation where two rules exert the same action  on a device under different conditions.  The environment conflict is akin to the opposite conflict and it describes the situation where two rules exert contrary actions on a shared device under different environmental conditions, e.g., ``If it is 5 AM, turn sprinkler ON''  vs. ``If it rains, turn sprinkler OFF''.

\subsubsection{Actuation Conflict on IoT Services}~\\
Recently, conflicts have been discussed in the context of IoT services. An IoT service can be defined by a set of functionalities and non-functional attributes.  The Actuation Conflict is termed as \emph{Functional Conflict} in  \cite{chaki2020conflict, chaki2020fine}. When multiple occupants simultaneously have distinct state requirements on a service's functional property, a functional conflict arises. For instance, one person might turn on the light while watching TV, and another resident might choose to turn it off. Multiple IoT services can be composed together and become a composite service \cite{huang2021conflict}. When multiple composite services run simultaneously, they may compete to access some shared devices or environment properties, leading to conflicts. The Function-Function conflict in \cite{huang2021conflict} can be mapped to our Actuation Conflict. It refers to the situation that two incompatible actions are exerted on a shared IoT service simultaneously \cite{huang2021conflict}. For example, turning on the light and turning off the light at the same is a \emph{Function-Function} conflict.

\subsubsection{Actuation Conflict on Home Appliances}~\\
Conflicts occur in the context of home networked service. An example of a home networked service is the ``Coming Home Light Service'' (i.e. when a door sensor notices that the user comes home, lights are automatically turned on. Then, the brightness of the lights is adjusted to an optimal value based on the current degree obtained from an illuminometer) \cite{nakamura2005feature}. A home networked service generally constitutes  features of multiple appliances. Two services may conflict with each other if they share  some appliances or environmental properties. This conflicting situation is referred to as feature interaction, including appliance interaction and environment interaction.  In particular, the appliance interaction can be mapped to our Actuation Conflict.  Appliance interaction is also  addressed in \cite{leelaprute2008detecting}, which refers to two services attempting conflict operations on the same appliance. For example,  the HVAC service and the air-cleaning service can operate the ventilator with conflicting purposes. Similarly, the appliance interaction is also discussed in \cite{nakamura2005feature}.  An \emph{appliance interaction} refers to the situation that two services simultaneously invoke   a common appliance in an incompatible manner. One subcategory describes that two services invoke incompatible methods of a common appliance at the same time, which can be mapped to our Actuation Conflict. For example, the TV cannot stay on the TV.Power(OFF) and   TV.Power(ON) states at the same time.

\subsection{Preference Conflict}
A body of works considers the Preference Conflict in the contexts of trigger-action rules \cite{ibrhim2020formal}, IoT apps  \cite{alhanahnah2020scalable}, policies \cite{nagendra2019viscr, shehata2007using}, IoT services \cite{chaki2020fine, chaki2020conflict}, and home appliances \cite{nakamura2005feature}.

\subsubsection{Preference Conflict on Trigger-Action Rules}~\\
A framework referred to as Rule Verification Framework is presented to detect and resolve conflicts between trigger-action rules \cite{ibrhim2020formal}. Among the seven types of conflicts discussed in this surveyed paper, the  \emph{Value Inconsistency Conflict (VIC)} can be generalized to our Preference Conflict category. The VIC refers to two rules that update the shared devices into different states by inputting different values.

\subsubsection{Preference  Conflict on IoT Apps}~\\
Various \emph{interaction threats} between apps are studied in \cite{alhanahnah2020scalable}. A behavior rule is first derived from the IoT app's source code  by linking the triggers, actions, and logical conditions of each control flow. In particular, the \emph{action-action (conflict) interaction threat} refers to the situation where two rules  act on the same device's attribute with different values\cite{alhanahnah2020scalable}. This type of interaction threat can be mapped to our Preference Conflict category.  

\subsubsection{Preference  Conflict on Policies}~\\
Various conflicts between policies are addressed in \cite{nagendra2019viscr}.  When two policies leave the IoT infrastructure in a state that is unstable or unpredictable in its behavior, a gap conflict occurs. The characteristics of the gap conflict suit the definition of our Preference Conflict and can be mapped to our Preference Conflict category. For example, a policy says that ``from 8 pm to 9 pm set the thermostat to 65F in bedrooms'' while another policy says that ``from 9 am to 9 pm set the thermostat to 74F''. The two policies are conflicting because they try to update the parameter of the  thermostat  into different values.
 
Feature interactions among policies are detected in \cite{shehata2007using}. A feature is defined as a system functionality built by developers. For example,  ``TV.ON()'' is  a feature for the TV.  A policy is defined as a piece of information that is specified by normal users and used to modify the behavior of a system through modifying features of the system. An example of a policy is  ``Open the windows between 3:30 pm and 4:40 pm''. Our Preference Conflict category can be used to map interactions between two system axiom policies. When the rule attribute of one system axiom simple policy conflicts with the rule attribute of a second system axiom simple policy, it describes how two system axiom simple policies interact. The fact that the policy ``Maintain the temperature of the hot water of the hot water tap in the kitchen to 45C" conflicts with the policy ``Maintain the temperature of the hot water of the hot-water-tap in the bathroom to 40C" is an illustration of the feature interaction.

\subsubsection{Preference  Conflict on IoT Services}~\\
Various conflicts of IoT services are considered in a multi-resident environment \cite{chaki2020fine, chaki2020conflict}.  The \emph{qualitative non-functional conflict} (resp. quantitative non-functional conflict) refers to a situation in that different people set nominal QoS preferences (resp. numeric QoS preferences)  over a single service at the same time duration and the same location. An example of a qualitative non-functional conflict is when two residents want  to watch different TV programs at the same time. An example of a \emph{quantitative non-functional conflict} is when two residents want to set the luminosity of a common light  to be  900\emph{lux}  and  600\emph{lux}  at the same time. Therefore, these two types of conflicts can be mapped to our Preference Conflict.

\subsubsection{Preference  Conflict on Home Appliances}~\\
One subcategory of the  \emph{appliance interactions} describes that two services input different values into the same method of a common appliance at the same time  \cite{nakamura2005feature}. It  can be mapped to our Preference Conflict. For example, the TV channel cannot be set into  [Channel==2] and [Channel==5.1 ] at the same time.

\subsection{State Impact Conflict} 
State Impact Conflict is widely considered in the majority of existing research in the context of trigger-action rules \cite{ibrhim2020formal,sun2014conflict,wang2019charting, zamudio2009understanding}, IoT apps \cite{alhanahnah2020scalable,celik2019iotguard,chi2020cross,li2020diac, shen2017systematic}, policies \cite{liu2019remediot,nagendra2019viscr,pradeep2021automating,shehata2007using}, and IoT devices \cite{alharithi2019detecting,ding2021iotsafe}.  

\subsubsection{State Impact Conflict on Trigger-Action Rules}~\\
The Rule Verification Framework is introduced to detect and resolve conflicts between trigger-action rules \cite{ibrhim2020formal}. Seven types of conflicts are discussed, among which the  \emph{ Rule Dependency Conflict (RDC)} can be generalized to our State Impact Conflict category. The RDC refers to the situation where the action generated by one rule leads to the triggering of another rule, whose action, in turn, triggers the first rule.  In  \cite{sun2014conflict}, the \emph{Direct Dependence Conflict} and \emph{Indirect Dependence Conflict} describe two specific scenarios of the State Impact Conflict in a smart building environment. Similarly, the \emph{Direct Dependence Conflict} describes the situation that the trigger condition of one rule is the state after executing another rule.   The \emph{Indirect Dependence Conflict} describes the situation in that multiple rules can be chained together via their trigger condition relations. The \emph{Indirect Dependence Conflict} for multiple rules can be inferred by pairwise \emph{Direct Dependence Conflict}. Six types of inter-rule vulnerabilities are addressed, among which the \emph{action loop} captures the characteristics of the State Impact Conflict \cite{wang2019charting}. The \emph{action loop} refers to the situation in which an object’s activation cyclically leads to its own re-activation. In \cite{zamudio2009understanding}, the dependency between rules in the connected devices may result in unwanted system behavior, which is rooted in the presence of a loop hidden among devices. A loop occurs when the state of one device is dependent on that of a second device. In this respect, the loop is similar to our State Impact Conflict.

\subsubsection{State Impact Conflict on IoT Apps}~\\
The \emph{action-trigger  }, \emph{action-condition (match)}, and \emph{action-condition (no match)} present three different scenarios of the State Impact Conflict occurring between IoT apps \cite{alhanahnah2020scalable}. Specifically, the action-trigger interaction threat refers to the situation where the action of a rule activates the triggers of another rule. The ``match action condition'' and ``no match action condition'' refers to the situation where the action of a rule enables/disables another rule's condition, respectively. An approach called IOTGUARD is developed to dynamically detect unsafe and insecure interactions between IoT apps   \cite{celik2019iotguard}. One type of interaction is that an app’s event handler might change device attributes and trigger events in another app. This type of interaction can be generalized to our State Impact Conflict.  For example, an app turns on the light in a room when smoke is detected, and another app locks the front door when the light is turned on. This type of interaction between apps potentially exposes the resident to the risk of a fire.

In \cite{chi2020cross}, both \emph{trigger-interference threats}   and  \emph{condition-interference threats} can be classified into the State Impact Conflict category. Specifically,  the trigger-interference threat  describes the situation where a rule's action  produces an event and thus triggers other rules.  For example, the  rule   ``If the voice command is received, then turn on the TV'' will accidentally trigger another rule   ``If the TV is turned on, then turn off the camera''. The condition-interference threat describes the situation where a rule's action enables/disables another rule's condition. For example, the rule ``If the user leaves home, then set the home to away-home mode'' would disable another rule `` If the door is opened and the home is sleep-mode, then turn on the burglar alarm''.   

App conflicts are also discussed in \cite{li2020diac, shen2017systematic}, i.e., \emph{strong conflicts} and \emph{weak conflicts}. The strong conflict can be mapped to our State Impact Conflict. A strong conflict refers to a situation where some actions of an app get disabled as an interaction with other apps. For instance, a home assistant app turns on the light whenever the door contact is open, while a security app begins recording video when the room is dark and a door contact is open. When the two apps run at the same time, the security app will not function normally as the home assistant app disables the video by preventing the needed condition from being met. 


\subsubsection{State Impact Conflict on Policies}~\\
Conflict among user-defined policies discussed in \cite{liu2019remediot}
is similar to our State Impact Conflict. A user-defined policy specifies the functionality of a trigger-action rule. For example, the user-defined policy ``user-away -> user-away-mode-on'' may conflict with another policy ``user-away-mode-on, temp-high -> windows-on'' because executing the former  may trigger the latter to be invoked. Similarly,  several types of conflicts are discussed, among which the  \emph{loops in automation conflict} suits the features of our State Impact Conflict \cite{nagendra2019viscr}. The loops in automation conflict can be loosely defined as a set of policies chained by triggers and actions. For example, a policy says that ``keep the main doors open between 6 pm to 10 pm'' while another policy says, ``To save energy, turn OFF the heating or cooling in that room if the primary doors and windows are left open for longer than five minutes.". A loop  in automation conflict arises as the opening main door triggered by the former policy would trigger the latter policy to execute. In this regard, the loop in automation conflict is aligned with the characteristics of our State Impact Conflict definition and can be mapped to our State Impact Conflict category.

The \emph{extended conflict} can be generalized to our State Impact Conflict category because it describes the state of a device as a partial trigger condition of another device \cite{pradeep2021automating}.  When a safety property is broken, a prolonged dispute results. An extended conflict with regard to a safety property, for instance, can be defined as opening and closing the door within a short period of time. For example,  the rule ``if temperature > 80F and door-close-time > 5 minutes, then open door'' conflicts with another rule ``If door-open-time > 3 minutes, then close door''. In addition, feature interactions among policies are detected in \cite{shehata2007using}.   A feature is defined as a system functionality built by developers. For example,  ``TV.ON()'' is  a feature for the TV.  A policy is defined as a piece of information that is specified by normal users and used to modify the behavior of a system through modifying features of the system. An example of a policy is  ``Open the windows between 5:00 pm and 6:30 pm''. The \emph{interactions between a dynamic behavior policy and a system axiom policy} is close to our State Impact Conflict as the execution of one policy disables another policy. It shows that when the action attribute of the dynamic behavior simple policy conflicts with the rule attribute of the system axiom simply policy, there is an interaction between the two simple policies.

\subsubsection{State Impact Conflict on IoT Devices}~\\
The \emph{indirect conflict} describes two specific situations (i.e., \emph{chain conflict} and \emph{feedback conflict})  of State Impact Conflict arising between rules \cite{alharithi2019detecting}. Chain conflict occurs when  the execution of one rule leads to the execution of another.   For instance, the first rule states, ``If Temperature > 80, open AC," and the second, ``If AC ON, close window."  Multiple chain conflicts can be used to deduce the feedback conflict. The three rules ``If the air purifier opens, then close the windows," ``If the windows are closed, then close the curtain," and ``If the curtain is closed, then open the air purifier'' are examples of feedback conflicts. A system called IoTSAFE is developed  to detect physical interactions of IoT devices \cite{ding2021iotsafe}. The State   Impact Conflict is termed as  the \emph{cyberspace interaction}. A cyberspace interaction describes the situation where apps interact through a common device or system event. For example, an app turns on the light automatically after sunset, while another app opens the door when the same light is turned on. These two apps act on the common device through the event of turning on the light.

\subsection{Environment Conflict}
Many works consider environment conflicts in the context of trigger-action rules \cite{sun2014conflict,xiao2019a3id,yu2021tapinspector}, IoT apps \cite{stevens2020comparing}, policies \cite{shehata2007using}, IoT devices \cite{chen2019multi,ding2018safety,ding2021iotsafe}, IoT services \cite{chaki2020fine, chaki2020conflict,huang2021conflict}, and home appliances \cite{leelaprute2008detecting, nakamura2005feature}.

\subsubsection{Environment Conflict on Trigger-Action Rules}~\\
The \emph{Environment Mutual Conflict} describes the situation that the impact on an environment entity caused by invoking one rule triggers another rule to execute \cite{sun2014conflict}. The \emph{Environment Mutual Conflict} can be mapped to our Direct Environment Impact Conflict.
For example, the rule ``temperature > 30C → open air-conditioner $\wedge$ keep temperature below 28C'' may conflict with another rule ``temperature < 25C → open heater $\wedge$  keep temperature
above 28C via the temperature channel. Implicit interference hidden among trigger-action rules is identified \cite{xiao2019a3id}.  Implicit interference refers to the situation that two or more rules simultaneously target multiple actuators that have contradictory effects on a shared environment property. An example of implicit interference is that one rule invokes the heater to heat the room while another rule invokes the air-conditioning to cool the same room. In this regard, such implicit interference can be mapped to our Indirect Environment Impact Conflict.   The \emph{Action-Trigger Interaction} is detected among rules \cite{yu2021tapinspector}. Action-Trigger Interaction refers to the situation where one rule   can activate  another rule directly or through physical channels such as humidity, temperature, and illumination. For example, the rule ``if the temperature drops below 10°C, turn heater on'' may have Action-Trigger Interaction with another rule ``if the temperature rises above 28°C, open window''. This is because turning on the heater can increase temperature and thus may activate the window to be opened. In this regard, the  Action-Trigger Interaction can be mapped to our Direct Environment Impact Conflict. 

\subsubsection{Environment Conflict on IoT Apps}~\\
 \emph{Multi-app coordination} between IoT apps are detected in a smart home system \cite{stevens2020comparing}. Multiple IoT apps are said to form the coordination if they can be chained by the postcondition/precondition of apps. In this regard, the multi-app coordination can be generalized to our Direct Environment Impact Conflict category.  Three IoT apps (i.e., MaliciousApp, HomeModeAPP, WindowControlApp ) are used  in the   example   to illustrate   the coordination.  Suppose the MaliciousApp is controlled by an attacker, and it sets the smart home into the home mode. The home mode triggers HomeModeApp to turn on the heater. The heating effect may trigger the WindowControlApp to open the window. The coordination of the apps makes the home   unsafe by opening the window when no one is at home.

\subsubsection{Environment Conflict on Policies}~\\
Feature interactions among policies are detected in \cite{shehata2007using}. The \emph{interactions between two dynamic behavior policies} can be mapped to our Indirect Environment Impact Conflict category. It specifically shows that two dynamic behavior simple policies interact when their previous state attributes and trigger event attributes are the same and action attributes are contradicting. For instance, when the temperature outside the house is lower than the temperature inside, the system opens the window while trying to heat the house.

\subsubsection{Environment Conflict on IoT Devices}~\\
The \emph{inter-app interaction} is extracted between IoT devices by the tool called IoT Interaction Extraction (IoTIE) \cite{chen2019multi}. The multiple apps form an \emph{inter-app interaction} if they are chained by physical environments. In this regard, we can generalize the inter-app interaction to our Direct Environment Impact Conflict category.  A framework called  IoTMon is presented to detect  \emph{inter-app interaction chains} that is much the same as aforementioned inter-app interaction \cite{ding2018safety} . Two apps form an inter-app interaction chain if they are connected by physical channels such as humidity, temperature, motion, and illumination. The inter-app interaction chain captures the key characteristics of our Direct Environment Impact Conflict. It describes the postcondition of using an IoT device impacts the environment property which makes the precondition of another IoT device come true and triggers this IoT device. Therefore, the inter-app interaction chain can be mapped to our Direct Environment Impact Conflict category. For example, the Heater Control App turns on the heater when it is after 18:00 while the Temperature Control App opens the window when the temperature is more than 85F \cite{ding2018safety}. In this regard, there is an inter-app interaction chain between the two apps because turning on the heater for a period of time may trigger the window to be opened accidentally.  
  
 In  \cite{ding2021iotsafe}, the \emph{physical interaction} is similar to our Environment Impact Conflict. A physical interaction describes the situation where an app impacts the physical environment properties such as temperature, humidity, and illumination. Apps that subscribe to several devices can communicate with one another through shared environment attributes thanks to such effects on the physical environment. For instance, when the temperature exceeds a user-defined threshold, one app turns on the heater while another app opens the windows. In this regard, the heater has a physical interaction with the window.  Therefore, the physical interaction can be mapped to our Direct Environment Impact Conflict category.
 
\subsubsection{Environment Conflict on IoT Services}~\\
The \emph{indirect service impact conflict} in \cite{chaki2020fine, chaki2020conflict} can be mapped to our Indirect Environment Impact Conflict. An indirect service impact conflict occurs when multiple residents prefer to use multiple services that have an impact on each other’s non-functional property. For example, a resident turns on the air-conditioning to cool the room while another resident opens the window to freshen the indoor air. An indirect service impact conflict may happen because opening the window may impact the air-conditioner's intended cooling effect.

Three types of environment impact conflict are detected in   \cite{huang2021conflict} namely, \emph{ Opposite-environment-impact-conflict, Cumulative-environment-impact-conflict}, and \emph{Transitive-environment-impact-conflict}.   Both Opposite-environment-impact-conflict and  Cumulative-environment-impact-conflict refer  to the situation that a shared environment property is changed  by two different IoT services. Thus, the two types of conflicts can be mapped to  our Indirect Environment Impact Conflict. The Transitive-environment-impact-conflict refers to the situation in which the usage of one IoT service may invoke another IoT service coincidentally due to their correlation via environment properties. This type of conflict can be mapped to our Direct Environment Impact Conflict.

\subsubsection{Environment Conflict on Home Appliances}~\\
The \emph{environment interaction } in \cite{leelaprute2008detecting, nakamura2005feature} suits the features of the Environment Impact Conflict. An environment interaction refers   to an indirect conflict among appliances via environment properties. Two subcategories of environment interactions are discussed. One subcategory of environment interactions describes the situation that two services update a shared environment property at the same time, which can be mapped to our Indirect Environment Impact Conflict. The other subcategory describes the situation that the impact on an environment caused by a service would invoke another service accidentally. This category of environment interactions can be mapped to our Direct Environment Impact Conflict.

{\scriptsize
\begin{longtable}{|p{.06\textwidth} | p{.12\textwidth}| p{.08\textwidth} | p{.08\textwidth} | p{.1\textwidth} | p{.14\textwidth} | p{.14\textwidth} | }
\caption{Classifying conflicts in related work using our conflict taxonomy}
 
\label{mapping} \\
 
\hline
 \textbf{References} & \textbf{Contexts} & \textbf{Actuation Conflict} &   \textbf{Preference Conflict} &   \textbf{State Impact Conflict} &  \textbf{Direct  Environment  Impact Conflict} &   \textbf{Indirect  Environment Impact Conflict} \\
\hline

\cite{ibrhim2020formal}&Trigger-action rules &\checkmark  & \checkmark &\checkmark   &   &    \\ 
\hline

\cite{sun2014conflict}&Trigger-action rules & \checkmark & & \checkmark  & \checkmark  &    \\ 
\hline

 \cite{wang2019charting}&Trigger-action rules & \checkmark  & &  \checkmark &  &   \\ 
\hline

\cite{xiao2019a3id}&Trigger-action rules &   &   &  &  &  \checkmark \\ 
\hline

\cite{yu2021tapinspector}&Trigger-action rules & \checkmark &  &    & \checkmark  &   \\ 
\hline

\cite{zamudio2009understanding}&Trigger-action rules &   &  & \checkmark    &  &   \\ 
\hline

\cite{alhanahnah2020scalable}&IoT apps & \checkmark & \checkmark & \checkmark  &   &   \\ 
\hline

\cite{celik2019iotguard}&IoT apps &\checkmark &  & \checkmark  &   &  \\ 
\hline


\cite{chi2020cross}&IoT apps &\checkmark& & \checkmark  &   &      \\ 
\hline

\cite{hamza2022hsas}&IoT apps & \checkmark &  &   &   &  \\ 
\hline

\cite{li2020diac}&IoT apps & \checkmark&  & \checkmark  &   &    \\ 
\hline

\cite{shen2017systematic}&IoT apps & \checkmark&  & \checkmark  &   &    \\ 
\hline

\cite{stevens2020comparing}&IoT apps &  &  &   &  \checkmark &    \\ 
\hline

\cite{trimananda2020understanding}&IoT apps &\checkmark &  &  &  &   \\ 
\hline

\cite{yagita2015application}&IoT apps &\checkmark &  &  &  &   \\ 
\hline
\cite{goynugur2017tractable}& Policies & \checkmark &  &  &   &   \\ 
\hline

\cite{liu2019remediot}& Policies & \checkmark &  & \checkmark  &   &   \\ 
\hline

\cite{nagendra2019viscr}&Policies &\checkmark & \checkmark & \checkmark  &   &    \\ 
\hline

\cite{pradeep2021automating}&Policies & \checkmark &  & \checkmark  &   &    \\ 
\hline

\cite{shehata2007using}&Policies &  & \checkmark & \checkmark  &   & \checkmark  \\ 
\hline



\cite{alharithi2019detecting}&IoT devices & \checkmark& &  \checkmark  &   &    \\ 
\hline 

\cite{chen2019multi}&IoT devices &  &  &   &  \checkmark &  \\ 
\hline

\cite{ding2018safety}&IoT devices &  &  &   & \checkmark  &     \\ 
\hline

\cite{ding2021iotsafe}&IoT devices &  &  & \checkmark  & \checkmark  &   \\ 
\hline

\cite{chaki2020fine}&IoT services & \checkmark &\checkmark  &   & \checkmark  & \checkmark  \\ 
\hline

\cite{chaki2020conflict}&IoT services & \checkmark &\checkmark  &   & \checkmark  & \checkmark   \\ 
\hline

\cite{huang2021conflict}& IoT services& \checkmark & &   & \checkmark & \checkmark  \\ 
\hline

\cite{leelaprute2008detecting} & Home appliances &  \checkmark &  &   & \checkmark  &  \checkmark  \\ 
\hline

\cite{nakamura2005feature}& Home appliances &\checkmark &\checkmark &   & \checkmark& \checkmark  \\ 
\hline

\cite{ahmed2021adaptation}& Home appliances &\checkmark &\checkmark &   & \checkmark &   \\ 
\hline

\end{longtable}
}

\section{Conflict  Detection} 
This section provides an overview of the conflict detection approaches in the existing works. We compare these approaches from different perspectives as shown in Table \ref{efficiency}. We first outline the evaluation protocols, then overview the most representative datasets. Finally, we elaborate on conflict detection approaches.

\subsection{Evaluation Protocol}
\begin{itemize}

    \item Categorize the approaches: We categorized the conflict detection approaches found in existing literature based on their main principles, methods, or algorithms, such as formal rule modeling approach, graph-based approach, object-oriented approach, model-checking approach, machine learning-based, etc.

    \item Identify the relevant parameters for each approach: For each approach, we identified key parameters, such as contexts (e.g., IoT devices, policies, trigger-action rules), main contributions (novel method or improvement over existing methods), domains (application areas), and data sources (where the information for conflict detection is derived from).

    \item Collect data for each parameter: We reviewed the selected publications and extracted the required information corresponding to each parameter in the table. This allowed us to systematically compare the conflict detection approaches based on the identified parameters.

    \item Populate the table: We compiled the extracted information in a table format, making it easy to visualize and compare the approaches based on the parameters.
    
\end{itemize}

\subsection{Datasets for Conflict Detection} 
\begin{figure*}[htbp]
\centering
\includegraphics[width=  \textwidth]{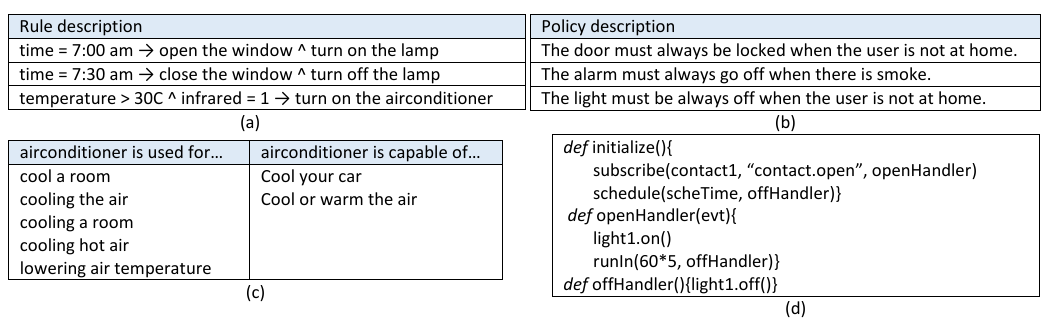}
\caption{(a)  Rule descriptions \cite{sun2014conflict}, (b)  Policy descriptions \cite{celik2019iotguard}, (c) Open-source knowledge about appliances \cite{xiao2019a3id}, and (d) IoT app source code \cite{li2020diac} }
\label{fig:datasource}
\end{figure*}

A variety of conflict detection approaches are designed according to diverse datasets. This section provides an overview of datasets used in the surveyed papers. We categorize datasets into four main groups, namely, app configuration code, rules, outsourced device knowledge, and policies. We provide a snapshot of each category of datasets and its associated representative references as shown in Fig.\ref{fig:datasource}. 

Several representative works focus   on extracting  triggers, conditions, and actions from the app configuration code \cite{li2020diac,shen2017systematic,chi2020cross,yu2021tapinspector}. These works generally perform  a static code analysis to extract  these three types of information before detecting conflicts. A representative example of app configuration code can be found in \cite{li2020diac,shen2017systematic}. Some works aim  to detect conflicts from policies \cite{celik2019iotguard}\cite{shehata2007using}\cite{nagendra2019viscr}. These policies are usually specified by residents or developers in plain text. Human knowledge is required to extract triggers, conditions, and action information from text-based policies before modeling and detecting conflicts. An example of policy description is ``The door must always be locked when the user is not at home'' \cite{celik2019iotguard}. Triggers, conditions, and actions can also be parsed from XML files and then assembled as rules. Conflicts are modeled and detected based on these rules \cite{sun2014conflict, magill2016exploring}. An example of the rule description ``time = 7:00 am $\to$ open window $\wedge$ open lamp'' \cite{sun2014conflict}. Some works model the environmental interactions between appliances by referring to open-source knowledge \cite{xiao2019a3id} \cite{huang2021conflict}. Open-source knowledge such as the ConceptNet\footnote{https://conceptnet.io/} provides descriptions of the appliance's properties, such as functionalities and environmental impact. For example, the air-conditioner can be described by the set of properties provided by different people ``< cooling a room, cool down, cool the environment, cool your home, cool a house, cooling,...>'' \cite{xiao2019a3id}. NLP techniques are usually applied to extract these properties. Conflicts are modeled and detected using a knowledge graph. Some works employ an object-oriented modeling approach to describe an appliance's properties \cite{alfakeeh2016feature}\cite{nakamura2005feature}\cite{nakamura2013considering}.  Each appliance is modeled by a set of methods and properties. For example, an air-conditioner unit is modeled by the set of methods ``<setPower(onoff), setTemperature(temp)>'', the set of pre-conditions ``<Power = `ON'>'', and the set of post-conditions ``<Power = onoff, TemperatureSetting = temp>''\cite{nakamura2005feature}.  Conflicts are detected by pairwise comparison between appliances. Online conflict detection requires the modeling of spatio-temporal aspects of IoT devices. Temporal information can be obtained by executing IoT devices, which are represented by device execution events.  There are two  representative types of device execution events, namely, time-point-based events and time-interval-based events. An example of a time-point-based event is `` Turning on switches at 4:49:16'' \cite{ding2018safety}. An example of a time-interval-based event is ``< Light switch, [4:30:23, 4:33:39]>''\cite{huang2021conflict}. Representative  online conflict detection based on time-point-based events and time-interval-based events are presented in \cite{ding2018safety}\cite{pradeep2021automating}  and\cite{chaki2020fine} \cite{huang2021conflict}, respectively.

\subsection{Conflict Detection Approaches} 
We  classify conflict detection approaches into five main categories: (i) graph-based approach, (ii) object-oriented approach, (iii) formal rule modeling approach, (iv) model checking approach, and (v) other approaches. For each related reference, we summarize its main contributions and identify application domains and collect data sources. 

\subsubsection{Graph-based approach}~\\
A   control flow graph approach named IoTCom  is proposed   to automatically detect interaction threats between IoT apps including (i) action-trigger (c.f., State Impact Conflict), (ii) action-condition-match  (c.f., State Impact Conflict), (iii) action-condition-no match  (c.f., State Impact Conflict), (iv) self coordination, (v) action-action-conflict (c.f., Actuation Conflict), (vi) action-action-repeat (c.f., Preference Conflict), and vii) exclusive event coordination (c.f., Actuation Conflict) \cite{alhanahnah2020scalable}. IoTCom constitutes two main components, i.e., \emph{behavioral rule extractor} and \emph{formal analyzer}. The behavioral rule extractor aims to extract a control flow graph for each IoT app through static code analysis. A behavioral rule graph is generated by combining the control flow graph and its associated conditions and  events. The formal analyzer aims to detect interaction threats. It relies on  the behavior rule graph, a base model that defines behavior rules for IoT apps  and assertions for safety and security properties. A tool called IoT Interaction Extraction(IoTIE)  is developed to extract the inter-app interaction (c.f., Direct Environment
Impact Conflict)  that is chained by physical environments shared by IoT devices \cite{chen2019multi}. There are 13 types of physical environments included in  \cite{chen2019multi} such as temperature, humidity, voice, and color. The tool IoTIE has four components. The first component focuses on collecting apps from   Samsung’s SmartThings   and IFTTT  platforms. The second component aims to construct  the dependency between event, trigger, and conditions, resulting in intra-interactions. This can be achieved by using static program analysis (i.e., The methods of \emph{definition}, \emph{preferences}, \emph{installed}, and \emph{update} are performed against program analysis ). The first component also aims to extract physical environment properties by analyzing application description information such as the application name, description, and annotations. As a result, the relations between applications and physical environments are constructed as a graph. The  third component focuses on identifying inter-app interactions based on the generated graph by verifying whether two rules share the same physical environment properties. Finally,    two strategies are devised to  mitigate the high-risk interactions by detecting   whether the installed applications and the to-be-installed applications can form insecure interactions and inform developers to pay attention to the interaction problem. 

A control flow graph-based approach is presented to detect cross-app interference threats from trigger-action rules \cite{chi2020cross}. Three categories of interference threats are discussed including action-interference threats (c.f.,  Actuation Conflict), trigger-interference threats (c.f., State Impact Conflict), and condition-interference threats (c.f., State Impact Conflict). The control flow-based graph is constructed by analyzing the configuration code of IoT apps. In particular, a symbolic executor is used to extract rule semantics in terms of triggers, conditions, and actions. These semantics are assembled into a control flow-based graph. The control flow-based graph  serves as a vehicle to detect interference threats via checking against the Satisfiability Modulo Theories (SMT) model. A risk ranking schema is designed to evaluate the seriousness of interference threats based on various attributes, including app functionalities, device attributes, and threat types. The risk ranking schema labels the   risk level of detected threats into high, medium, and low categories. The IoTMon framework is introduced to detect  inter-app interaction chains (c.f., Direct Environment
Impact Conflict) using a graph-based approach \cite{ding2018safety}. The   framework  IoTMon constitutes three components, i.e., Application Analysis, Interaction Chain Discovery, and Risk Analysis. The Application Analysis component aims to identify the trigger-action control dependency of apps and physical channels. The trigger-action control dependency can be extracted through static program analysis. The physical channels can be identified by analyzing app description information provided by developers. The identified trigger-action control dependency and physical channels serve as input into the Interaction Chain Discovery component, which uses a graph to combine the two types of information. Inter-app interaction chains are identified if   trigger-action control dependency is connected through physical channels. The  Risk Analysis component aims to evaluate the risk level for the discovered inter-app interaction chains. A risk evaluation mechanism is designed  based on the computation of distances between the inter-app interaction chains and the baseline. A large distance represents a high-risk level and should be properly resolved. 

A system called IoTSAFE is developed to detect physical interactions of IoT devices including cyberspace interactions (c.f., State
Impact Conflict) and physical interactions (c.f., Direct Environment Impact Conflict)  \cite{ding2021iotsafe}. The  system IoTSAFE aims to identify and predict physical interactions considering contextual features in   smart homes. In particular, the IoTSAFE system presents a runtime physical interaction discovery approach. This approach uses  static code analysis for constructing an interaction graph. Dynamic testing techniques are used to discover runtime physical interactions among IoT devices. The interaction graph, combined with the temporal physical interaction graph, serves as a vehicle for predicting future risky situations and disabling unsafe device states. This paper focuses on only one type of conflict and ignores other types of conflicts. The strength of this paper is that it provides an interaction prediction mechanism. A framework REMEDIoT is designed to detect and resolve conflicts in IoT-based smart environments \cite{liu2019remediot}.  Two categories of conflicts are considered, i.e., racing event (c.f., Actuation Conflict) and cyclic event (c.f., State Impact Conflict). It detects conflicts in a given set of IoT applications with respect to a set of policies, i.e., rules that define allowable and restricted state-space transitions of devices. In detail, an \emph{actuation graph} is provided to abstract actuators among IoT applications. The graph is a directed graph where each node represents an actuation model that hides the implementation units for developers. The direction of the graph stands for the dependency relation among actuation modules. REMEDIoT checks if a given command triggers any conflicts using the actuation graph.

A knowledge graph-based approach is proposed to explore the direct and indirect conflicts among IoT services \cite{huang2021conflict}. Specifically, four types of conflicts are discussed, namely, Function-function conflict (c.f., Actuation Conflict), Opposite-environment-impact-conflict (c.f., Indirect Environment Impact Conflict),
Cumulative-environment-impact-conflict (c.f., Indirect Environment Impact Conflict), and Transitive-environment-impact-conflict (c.f., Direct Environment Impact Conflict). The conflict detection approach relies on a knowledge graph to capture the direct and indirect relations among IoT services.   Three types of relations, including the \emph{increase, decrease, trigger} relations, are  considered and extracted automatically using NLP techniques. The conflict detection approach can identify possible conflicts in a real-time fashion (online) from IoT service event sequences.  The obvious advantage of the knowledge graph-based approach is that it extracts the direct and indirect relations among IoT services automatically and requires few human annotations. The knowledge graph for IoT services is generic and can be tailored to a new smart home environment based on its context. The limitation of this approach is that the detected conflicts may be redundant and can be ignored because the closed scope of effect of an IoT service is not considered (e.g.,  air-conditioning and kettle may be considered as an Opposite-environment-conflict). An approach called VISCR is introduced to detect a variety of conflicts that arise between policies \cite{nagendra2019viscr}. Several
types of conflicts are discussed including gap conflict (c.f., Preference Conflict), loops in automation conflict (c.f., State Impact Conflict), and potential conflict (c.f., Actuation Conflict). The  approach VISCR uses a vendor-independent graph-based policy specification mechanism to translate policies into a graph because the policies are specified by different programming languages. For the purpose of detecting conflicts, these vendor-independent graph-based policies are provided as input to the vendor-independent conflict detection and resolution engine (VICE). The detected conflicts are sent to an engine to automatically resolve   conflicts.   Unresolved conflicts are reported to the stakeholders for manual resolution. The time complexity of the proposed algorithm is $O(L* L_s )$ where $L$ is the list of policies and $L_s$ is the number of graph nodes.

A graph-based approach is introduced to detect strong conflicts (c.f.,  State Impact Conflict) and weak conflicts (c.f., Actuation Conflict) among apps in an open IoT system \cite{li2020diac}. A graph structure named \emph{IA Graph} is used to describe the controls in each app and the various schedules of events. Based on the  \emph{IA Graph}, an efficient algorithm is designed to detect both strong and weak conflicts using first-order logic. This paper also proposes an innovative conflict resolution policy considering seriousness levels of conflicts, conflict frequency, and users' preferences and interests. An information flow graph-based approach called IRULER is presented to detect inter-rule vulnerabilities hidden among trigger-action rules \cite{wang2019charting}. Six types of inter-rule vulnerabilities are discussed including condition bypass (c.f., Actuation Conflict), condition
block (i.e. it is not captured by our taxonomy), action revert (c.f.,  Actuation Conflict), action conflict (c.f.,  Actuation Conflict), and action loop (c.f., State Impact Conflict). The inter-rule vulnerability is modeled by an information flow graph and detected by inspecting the text descriptions of triggers and actions using NLP techniques. In \cite{shen2017systematic},   strong inter-app conflicts (c.f., State Impact Conflict)  and weak inter-app conflicts (c.f., Actuation Conflict) are studied in smart homes. A new method is proposed that   the transitions inside an app are represented as Satisfiability Modulo Theories (SMT) formula. An SMT solver is then used to find conflicts between different programs. The benefit of utilizing an SMT solver is to create model representations that precisely depict how the device controls interaction in an application's source code. The approach only takes into account simple logic, which includes changing a device's state with a binary value, i.e., on or off, and ignores complex application logic found in condition statements with numerous time and threshold values.

A knowledge graph-based method is presented  to detect implicit interference (c.f., Indirect Environment Impact Conflict) hidden among trigger-action rules \cite{xiao2019a3id}. The  approach A3ID (Automatic and Interpretable Implicit Interference Detection), is to identify conflicts automatically based on the knowledge graph. A3ID extracts description information regarding each device from the publicly available knowledge graph ConceptNet. Then, A3ID uses NLP techniques to analyze the effect scope of each device (i.e., Scope Analysis), identify affected environment properties (i.e., Effect Evaluation), and analyze the interference relationship such as ``warming vs. cooling'' (i.e., Polarity Calculation). This approach also provides an explanation for detected conflicts. The noticeable strength of the A3ID approach is that it can detect conflicts automatically with few human annotations for devices. It also provides a method to organize devices  properly for conflict interpretation. The limitation of this approach is that it considers only one type of conflict and does not consider other types of conflicts. It also highly relies on programmers to input parameters in computing the interference relations between device and environment properties, which may produce unwanted interference relations. A graph-based approach is used to detect loops (c.f., State Impact
Conflict) between rules  \cite{zamudio2009understanding}. More specifically, a directed graph is adopted to model the interaction networks between devices, and loops are detected.

\subsubsection{Object-oriented approach}~\\
An object-oriented approach is presented to model networked home appliances for detecting conflicts (i.e., also referred to as feature interaction ) \cite{nakamura2005feature}. Two categories of conflicts are considered including appliance interactions
and environment interactions. Appliance interactions have Actuation Conflict and Preference Conflict sub-types and environment interactions have Indirect Environment
Impact Conflict and   Direct Environment Impact Conflict sub-types. An offline feature interaction detection approach is proposed in \cite{nakamura2005feature} based on the object-oriented model for home appliances and environment entities. Specifically, home appliances and environment entities are modeled as a set of methods associated with a set of properties. For example, an air-conditioning unit has a set of features such as AC.ON() and AC.setTemperature().  Feature interactions (i.e., appliance interactions and environment interactions) are identified by checking pairwise methods and validating by humans. 

The object-oriented approach introduced in \cite{nakamura2005feature} is further improved in \cite{igaki2010modeling} by considering environment property during feature interaction detection. Conflicts in this context are described as interactions between the appliance. If there is a direct conflict between two methods of the same appliance device, the appliance interaction is in conflict; however, if there is a conflict between two ways of distinct appliances that update the same environmental object, the conflict is indirect. This model is further improved in \cite{nakamura2013considering} by taking into account both environmental requirements and their corresponding impacts. An environment impact model is introduced to quantify how much an  environment property changes contributed by using an appliance. An environment requirement model is introduced to describe the user's expected environment state.  The  improved appliance interaction model and environment interaction model are   capable of capturing more conflicts that may be ignored by previous approaches.  

\subsubsection{Formal rule modeling approach}~\\
A framework, called Rule Verification Framework, is presented to address the issue of detecting and resolving conflicts between trigger-action rules \cite{ibrhim2020formal}. Seven types of conflicts are discussed including Rule Dependency Conflict (c.f., State Impact Conflict), Redundancy Conflict (c.f., Actuation
Conflict), Condition Inconsistency Conflict (c.f.,  Actuation
Conflict), Non-Specified Conflict (c.f.,   Actuation
Conflict), Value Inconsistency Conflict (c.f., Preference Conflict),  Useless Rule Conflict (c.f., N.A), and Range Incomplete Conflict (c.f., N.A). The   Rule Verification framework is built based on state-of-the-art Satisfiability Modulo Theories and uses a model checking method to accurately detect conflicts.  Time and location information is used to filter out rules for speeding up the algorithm. The framework also presents two strategies for conflict resolution. The first strategy is prioritization which relies on residents to assign priority to each rule to represent the relative importance of the rule. The second strategy requires residents to update rules' conditions or actions. Both conflict resolution strategies require extensive human involvement and lack applicability to other smart homes. A formal modeling approach is presented to identify conflicts between IoT systems dynamically and proactively resolve conflicts \cite{pradeep2021automating}. Two types of conflicts are considered, i.e., direct conflict (c.f., Actuation Conflict) and extended conflict (c.f., State Impact Conflict). To formalize and detect conflicts, the IoT system behavior is modeled by a set of operational policies and a set of safety properties. An operation policy describes a subsystem moving from one state to another state, while a safety property specifies the constraints on the behavior of the system. Both operational policies and safe properties are defined as a triplet <precondition, action, post-condition>. An example of  an operation policy is ((temperature > 25C and cooler = off), turn on cooler, cooler = on). An example of a safety property is (cooler = on and heater=on, turn off the heater, heater = off). Both the operation policy and safety property can be converted into first-order logic expressions. Therefore, the problem of conflict detection is transformed into checking the Boolean satisfiability problem. Linear Temporal Logic is combined with first-order logic to express temporal properties in terms of discrete-time slots. The NuXMV model checker tool is used to detect conflicts. NuXMV's output either shows satisfiability (i.e., no conflict) or a collection of counterexamples that can be applied to resolve conflicts.

 A semi-formal modeling approach, called IRIS (Identifying Requirements Interactions using Semi-formal methods), is developed to detect feature interactions among policies in a smart home environment \cite{shehata2007using}. Three main categories of feature interactions are considered, i.e., (i)
interactions between two system axiom policies (c.f., Preference Conflict). (ii) interactions between a dynamic behavior policy and a system axiom policy (c.f.,  State Impact Conflict), and (iii) interactions between two dynamic behavior policies (c.f., Indirect Environment Impact Conflict). IRIS is a six-step procedure that generates graphs and tables in the first five steps and detects feature interactions in the last step. IRIS first classifies policies into two categories, i.e., System Axiom policies and Dynamic Behavior policies. Then, it identifies various attributes for policies, extracts trigger events, identifies linked events, and generates graphic representations for each policy.  Finally, it applies the three aforementioned feature interactions as guidelines to detect hidden interactions by pairwise comparing graph-represented policies. It highly relies on developers to determine the feature interaction. A formal rule modeling approach is introduced to identify conflicts among trigger-action rules \cite{sun2014conflict}. A formal rule model UTEA based on User, Triggers, Environment entities, and Actuators is proposed to represent  rules. Then, the formal rule model is used to define 11 types of rule relations. Based on the 11 types of rule relations, five categories of conflicts among rules are modeled and detected.

\subsubsection{Model checking approach}~\\
A model checking-based approach is presented to detect various categories of conflicts hidden among trigger-action rules in  \cite{alharithi2019detecting}. Both direct and indirect conflicts are discussed and defined conceptually. Direct conflicts (c.f., Actuation Conflict) are further divided into three subcategories including opposite conflict, overwrite conflict and environmental conflict.  
Indirect conflicts (c.f., State Impact Conflict) have chain conflict  and feedback conflict sub-types. A model checking technique is proposed to detect different types of conflicts manually. The conflict detection model is designed from device, value, and action perspectives. Two rules are checked if they have the same/different devices, same/different values, and same/different actions and are categorized into the aforementioned conflict types. A  policy-based behavioral enforcement system, IOTGUARD, is presented to detect dynamic interactions between apps \cite{celik2019iotguard}. Three types of insecure interactions are discussed, i.e., (i).  interactions  that an app’s event handler might change device attributes and trigger events in another app (c.f., State Impact Conflict), (ii). interactions   that multiple apps can change the same attribute of a device at the same time (c.f., Actuation Conflict), and  (iii) interactions that apps could respond to the same event in conflicting ways (c.f., Actuation Conflict). Basically, IOTGUARD checks the app against a set of  safety and security properties before executing this app. Three key components are required  by IOTGUARD, a code instrument, a data collector, and   a security service. The code instrument  aims to collect the app's run time information, including  events, actions, and predicates. The collected information is used by the data collector to  describe the dynamic behavior of apps based on a state machine which is defined by states and state transitions. When the data collector receives an event, a corresponding action is requested. The security service evaluates the requested action against a set of IoT safety and security properties on the aforementioned dynamic behavior model by means of reachability analysis. If the requested action fails to pass a safety or security property, the security service rejects the requested action or presents an interface to users for approval.

A hybrid security system  HSAS-MD analyzer is developed to accurately detect action conflict (c.f.,  Actuation Conflict) between rules   \cite{hamza2022hsas}. It combines the model checking techniques and the deep learning-based approach. The   framework constitutes three main phases, including hybrid analysis,   deep-learning CNN modeling, and   model checking. The hybrid analysis component aims to perform data pre-processing by transforming the source code of IoT apps into the required format. It performs   static and dynamic analysis. The static analysis aims to convert source code into trigger-action rules. The dynamic analysis aims to extract control flow graphs from the source code. In the second phase, the   deep-learning CNN model aims to filter our fake actions before model checking. The first phase's trigger action rules are entered into the CNN model, which depicts them as a word-embedding vector. Suppose the trigger vector and condition vector are identical to those that have already been restored and used in the application under consideration. Action must correspond to new action. It is a benign action if the two activities line up. Otherwise, it is classified as malware. The third phase is model checking.  It verifies the static and dynamic features with the properties that describe all specifications of this app. 

A Linear Temporal Logic (LTL) based model checker, SPIN, is introduced to detect feature interactions in the home network systems considering Actuation Conflict, Direct and Indirect Environment
Impact Conflict \cite{leelaprute2008detecting}.  One of the key capabilities of the home network system is to integrate different features of appliances for providing value-added services. An example of such a service is the  Air-cleaning service that turns on the ventilator automatically when the smoke sensor senses smoke. Concurrent execution of these services may cause unexpected behaviors of the system, even when each service is independently correct. This situation is referred to as feature interaction. To detect the feature interactions, the home system  is modeled as three components: the environment, appliances, and services that are specified in the Promela language. Feature interactions with services, appliances, and the environment are formalized accordingly. LTL formulas are used to represent the correct properties that the system must hold. The SPIN model checker determines whether or not the specified property holds when the Promela code and an LTL formula are provided. In the event that the property is false, SPIN generates a counterexample. This counterexample makes it simple to identify the origin of the property violation.

An empirical study is conducted on conflicts among IoT-based applications \cite{trimananda2020understanding}. Two categories of conflicts are discussed including feature conflicts (c.f.,  Actuation Conflict) and saved-state conflicts (c.f., Preference Conflict). A model checking approach is proposed to identify conflicts from the code perspective of the IoT-based applications. A tool IoTCheck is developed to automatically identifies conflicts by model checking pairs of apps. Specifically, the state machine is used for  conflict detection in \cite{trimananda2020understanding}. A state machine is defined by a set of nodes  that represent the IoT device states and edges  that represent transitions between  states. Each transition  is associated with a corresponding sequence of actions. In \cite{yagita2015application}, a system is developed to provide install-time conflict (c.f., Actuation Conflict) detection using a model checking approach. The SPIN model is used to check the assertion ``no two apps use actuators to create different effects at the same location''. It also supports users in resolving conflicts by prioritizing apps and dividing them into groups of the same situation. A system called TAPInspector is developed   to detect rule interactions in concurrent trigger-action programming-based systems using a model checking approach \cite{yu2021tapinspector}. Several types of rule interactions are discussed including Action-Trigger
Interaction (c.f., Direct Environment Impact
Conflict), Action Duplication (c.f., Actuation Conflict), Action Conflict (c.f., Actuation Conflict), Action Overriding (c.f., Actuation Conflict), Action Breaking (c.f., N.A), Condition Blocking (c.f., r State Impact Conflict category), and Device
Disabling (c.f., N.A). The   system  TAPInspector first automatically extracts trigger-action  rules from IoT apps. Then it transforms these rules into a  hybrid model with model slicing and state compression. Taking the hybrid model and rule interaction models as inputs, it performs model checking against various safety and temporal properties.


\subsubsection{Other approaches}~\\ 
To identify conflicts of IoT services in a multi-tenant residential context, a fine-grained conflict detection framework is presented \cite{chaki2020fine}. The suggested conflict detection framework's architecture is made up of three parts: service event sequences, service usage patterns, and a conflict detection approach are three examples. The history of the inhabitants' service consumption is represented through service event sequences. From the prior usage patterns, a model of service usage habits has been developed. The usage habit model, which is based on consistency score, introduces the idea of Fuzzy Service Attribute. The conflict detection module then receives the habit model to identify various conflict types. Using the ideas of information entropy and information gain from the field of information theory, the dissimilarity score between the habits of various inhabitants is determined. Based on each instance's consistency score, the information entropy gauges how disorderly it is, and the information gain gauges how sparsely distributed those consistency scores are. The link between information entropy and information gain is used in the proposed approach to categorize three different forms of conflicts. They are Strong Conflict (c.f., Actuation Conflict), Tau Conflict (c.f., Preference Conflict), and Weak Conflict (c.f., Environment Impact Conflict), respectively. It is considered a serious conflict when locals use IoT services in highly disparate ways. When the difference is moderate, it is referred to as a tau dispute (i.e., leaning towards conflict). Weak conflict is indicated when there is only a very slight difference (i.e., when service usage habits are practically identical). To identify conflict before it occurs, the suggested conflict detection method uses temporal proximity techniques (i.e., apriori conflict detection). The dataset has been pre-processed using some complex statistical techniques, including binning and value stabilization.

A  hybrid approach (i.e., using both knowledge-driven and data-driven) is presented   to detect a variety of conflicts of IoT services \cite{chaki2020conflict}. Each IoT service is defined by a set of functionalities and non-functional properties. Conflict on individual IoT service is categorized as Functional Conflict
and Non-functional Conflict. Non-functional conflict is further sub-divided into three categories and they are Resource Capacity Conflict (c.f., N.A), Qualitative Non-functional Conflict (c.f., Actuation Conflict), and Quantitative Non-functional Conflict (c.f., Preference Conflict). On the other hand, conflict on multiple IoT services is categorized as Direct Service Impact Conflict (c.f., Direct Environment Impact Conflict) and Indirect Service Impact Conflict (c.f., Indirect Environment Impact Conflict). The proposed system has   IoT Service Conflict Ontology, IoT Service Usage History, and Conflict Detection components.   From past interaction data, the first component extracts inhabitants' preferences for using services. Conflict ontology, the second component, provides explicit and formal descriptions of various conflict types based on both functional and non-functional aspects of IoT services. The third element, the conflict identification algorithm, determines the type of conflicts by comparing ontology and usage data. The method is based on time interval overlap and frequent service use.  Likewise, an ontology-based approach is used to express policies and reason about policy conflicts (c.f., Actuation Conflict) \cite{goynugur2017tractable}. The policy is formalized using OWL-QL language which  supports efficient reasoning mechanisms and can be enhanced using the power of relational database query answering in the backend.  Policy conflicts are then detected using ontological reasoning by checking three conditions:  (a) policies should be applied to the same device, (b) one policy must oblige an action, while the other prohibits the same action; and (c) policies are active at the same time.

Two approaches are described to identify the multi-app coordination (c.f., Direct Environment
Impact Conflict), namely, the state-based approach and the rule-based approach   \cite{stevens2020comparing}. An empirical comparison study shows that the rule-based approach is more scalable than the state-based approach. Basically, the state-based approach models the   system by a tuple which constitutes  a set of states, a set of initial states, a transition relation matrix, and a labeling function.  In this regard, the smart home system can be described by a sequence of states which is denoted as device-attribute-value triplets. A list of states and labels is used to specifically characterize the smart home system. IoT devices, the attributes they are coupled with, and the possible values for each such attribute determine each state. Finding all feasible routes to reach a particular state is how the challenge of identifying multi-app coordination is transformed given a set of device-attribute-value triplets. Such multi-app coordination is found using the Alloy technique, which is based on the state-based model. In the rule-based method, the system is represented by a graph with a set of vertices and edges. In the smart home system, each vertex corresponds to a rule or an IoT app. If the triggers, conditions, and/or actions of two vertices use the same devices, attributes, or values, then those two vertices are connected by an edge. Two vertices can be joined to determine if there is multi-app cooperation.

{\scriptsize
\begin{longtable}{|p{.06\textwidth}|p{.15\textwidth}|p{.09\textwidth}|p{.30\textwidth}|p{.10\textwidth}|p{.15\textwidth}|}
\caption{Comparison of conflict detection approaches in existing literature}
\label{efficiency} \\
\hline
\textbf{References} & \textbf{Approaches} & \textbf{Contexts} & \textbf{Main Contributions} & \textbf{Domains} & \textbf{Data Sources}   \\\hline

\cite{alhanahnah2020scalable}   &
Graph-based approach & 
IoT apps    & 
An   approach  called  IoTCom is proposed to extract a control flow graph for each IoT app through static code analysis.  Multiple control flow graphs are assembled as a behavior graph which is used for detecting inter-app threats. & 
IoT system   & 
SmartThings apps and   IFTTT applets  
\\ \hline 

\cite{chen2019multi}  &
Graph-based approach & 
 IoT devices   & 
A tool called IoT Interaction Extraction (IoTIE)  is developed to extract interactions between IoT devices.  & 
IoT system  & 
SmartThings apps and   IFTTT applets 
\\ \hline

\cite{chi2020cross}   &
Graph-based approach  & 
 IoT apps & 
A control flow graph-based approach is proposed to detect cross-app interference threats from trigger-action rules by static code analysis. & 
Smart Home  & 
SmartThings apps
\\ \hline

\cite{ding2018safety}  &
 Graph-based approach & 
 IoT devices & 
A framework called IoTMon is proposed to detect inter-app interaction chains using a graph-based approach.  & 
IoT system  & 
SmartThings apps
\\ \hline

\cite{ding2021iotsafe}  &
Graph-based  approach  & 
 IoT devices & 
A system called IoTSAFE is developed to detect physical interactions of IoT devices using static code analysis. & 
IoT system  & 
SmartThings apps 
\\ \hline

\cite{liu2019remediot}  &
 Graph-based approach& 
 Policies& 
A remedial action framework for detecting and resolving IoT conflicts. & 
Smart environment    & 
SmartThings apps and IFTTT applets
\\ \hline

\cite{huang2021conflict} & 
Graph-based approach  & 
 IoT services & A knowledge graph, and a conflict taxonomy are used to detect conflicts dynamically. & 
Smart home   & 
CASAS dataset and synthetic dataset 
\\ \hline

\cite{nagendra2019viscr}  &
Graph-based approach & 
 Policies  & 
An approach called VISCR is presented to detect a variety of conflicts that arise between policies. & 
Smart building  & 
Synthetic policies and SmartThings apps
\\ \hline

\cite{li2020diac}   &
Graph-based approach    & 
  IoT apps & 
A graph-based approach, called IA Graph, is introduced to describe the controls in each app and the various schedules of events. Based on the IA Graph, an efficient algorithm is designed to detect both strong and weak conflicts using first-order logic. & 
IoT system    & 
SmartThings apps 
\\ \hline

\cite{shen2017systematic}   &
 Graph-based approach   & 
 IoT apps  & 
A system called DIAmond, is developed to detect inter-app conflicts. An efficient algorithm is designed using first-order logic to detect conflicts. & 
IoT system    & 
SmartThings apps 
\\ \hline

\cite{wang2019charting} & 
Graph-based approach  & 
 Trigger-action rules & 
A framework \emph{IRULER} is designed to detect inter-rule vulnerabilities. The inter-rule vulnerability is modeled by an information flow graph
and detected by inspecting the text descriptions of triggers and actions using NLP techniques. & 
 Smart home &  IFTTT applets 
 \\ \hline

\cite{xiao2019a3id}   & 
Graph-based approach   & 
 Trigger-action rules& 
A knowledge graph is learned using NLP techniques and used for inferring conflicts among rules. & 
Smart home   & 
IFTTT applets 
\\ \hline

\cite{zamudio2009understanding}   & 
Graph-based approach   & 
 Trigger-action rules& 
A graph-based approach is devised to detect loops between rules. More specifically, a directed graph is proposed to model
the interaction networks between devices and loops are detected. & 
Smart home   & 
Synthetic datasets
\\ \hline

\cite{nakamura2005feature}  & 
Object-oriented approach  & 
 Home appliances & 
Home appliances and environment entities are modeled by methods and properties based on which appliance interactions and environment interactions are identified. & 
Home network system  & 
Synthetic services
\\ \hline

\cite{igaki2010modeling}  & 
Object-oriented approach  & 
Home appliances  & 
Home appliances and environment properties are modeled by methods and properties. Appliance and environment interactions are defined based on methods and properties. & 
Home network system  & 
Synthetic services 
\\ \hline

\cite{nakamura2013considering}  & 
Object-oriented approach  & 
 Home appliances & 
An environment impact model and an environment requirement model are proposed for detecting appliance interactions and environment interactions based on the object-oriented approach. & 
Home network system    & 
Synthetic services 
\\ \hline  

\cite{ibrhim2020formal}  &
Formal rule modeling approach & 
 Trigger-action rules  & 
A framework is presented to detect and resolve conflicts between rules. A model checking method is applied to detect conflicts against user-defined policies. & 
Campus building   & 
Synthetic  services and   policies  
\\ \hline

\cite{pradeep2021automating}  &
Formal rule modeling approach & 
  Policies & 
A formal modeling approach is introduced to identify conflicts between IoT systems dynamically and proactively
resolve conflicts.   & 
IoT system  & 
Synthetic policies
\\ \hline

\cite{shehata2007using}   &
Semi-formal modeling approach  & 
  Policies& 
An approach called IRIS is devised to detect feature interactions among policies in a smart home environment based on an interaction taxonomy.  & 
Smart home   & 
Synthetic policies 
\\ \hline

\cite{sun2014conflict}  &
 Formal rule modeling approach   & 
 Trigger-action rules & 
A formal rule model (UTEA) is introduced based on Users, Triggers, Environment entities, and Actuators, based on which 11 types of rule relations are identified. Based on the rule relations, different types of rule conflicts are modeled and detected. & 
Smart building  & 
Rules 
\\ \hline

\cite{alharithi2019detecting}  &
Model checking  approach & 
  IoT devices & 
A conceptual framework is proposed to analyze conflicts from different perspectives. A model checking approach is used to detect conflicts based on the proposed conceptual framework.  & 
Smart home   & 
Synthetic rules
\\ \hline

\cite{celik2019iotguard}  &
Model checking approach & 
 IoT apps  & 
A  system, IOTGUARD, is developed to dynamically detect interactions between apps  by checking the app against a set of safety and security properties before executing this app. & 
IoT system    & 
SmartThings apps and   IFTTT applets 
\\ \hline

\cite{hamza2022hsas}  &
Deep Learning approach and model checking approach & 
IoT apps   & 
A framework, called HSAS-MD analyzer, is presented to detect malware by combining model
checking   and deep learning approaches. & 
Smart home   & 
SmartThings apps and IFTTT applets
\\ \hline

\cite{leelaprute2008detecting}  &
Model checking approach & 
 Home appliances& 
A feature interactions detection framework is proposed in the home network system using the SPIN model checker and PROMELA language. & 
Home network system   & 
Synthetic services
\\ \hline

\cite{trimananda2020understanding} &
Model checking approach   & 
 IoT apps & 
A tool called IoTCheck is developed to
automatically identify conflicts by model checking pairs of apps. & 
Smart home  & 
SmartThings apps 
\\ \hline

\cite{yagita2015application} &
Model checking approach   & 
 IoT apps & 
A system is developed to provide install-time
conflict detection using SPIN model-checking approach and resolve conflicts based on user priorities. & 
Smart home  & 
Apps 
\\ \hline

\cite{yu2021tapinspector}  &
Model checking approach & 
 Trigger-action rules  & 
A system called TAPInspector is developed to detect vulnerabilities in  IoT-based systems using a model checking approach based on static code analysis. & 
IoT system  & 
SmartThings apps and   IFTTT applets 
\\ \hline

\cite{chaki2020fine} &
Hybrid approach & 
IoT services    & 
A probabilistic conflict detection approach. Temporal proximity for IoT services is computed to evaluate the conflicting levels. & 
Smart home   & 
CASAS dataset and synthetic dataset
\\ \hline

\cite{chaki2020conflict} &
 Hybrid approach& 
 IoT services & 
A hybrid approach is designed to detect conflicts of IoT services by combining conflict ontology and statistical analysis techniques. & 
Smart home & 
CASAS dataset and synthetic dataset 
\\ \hline

\cite{stevens2020comparing}  &
State-based model and rule-based model & 
  IoT apps & 
A state-based approach and
  a rule-based approach   is proposed to identify the multi-app coordination. & 
Smart home   & 
SmartThings apps and IFTTT applets 
\\ \hline

\end{longtable}
}

\section{Open Issues and Future Directions}
We identify the following open problems which need to be addressed by further research. We focus on those issues that relate to conflicts and require substantial innovation to speed up the deployment of IoT-based smart homes. 

\textbf{Scalability of a conflict model:}
Conflict modeling is considered a prerequisite for detecting conflicts accurately and completely. However, most existing work focuses on some types of conflicts and ignores others. To the best of our knowledge, not much work can capture all conflicts discussed in our conflict taxonomy. As the evidence in Table \ref{mapping},  we can see that most of the work can capture the Actuation Conflict and State Impact Conflict. A fewer work can capture the Preference Conflict and the Direct/Indirect Environment Impact Conflict. The reason behind the incomplete coverage of conflicts is that the existing conflict models ignore some important aspects, such as side-effects of IoT devices, spatio-temporal aspects, users' preferences, and priorities. Therefore, we believe a more scalable and extensible conflict model is needed to capture a broader range of conflicts considering various aspects of smart homes.
 
\textbf{Contexts in smart homes:} 
Any information that can be utilized to describe the status of an entity is considered context \cite{dey2001conceptual}. An entity, including the user and apps themselves, is any individual, location, or thing that is thought to be important to the interaction between a user and an application \cite{huang2022multi}. Most existing works on conflict modeling and detection mainly consider environmental properties. However, other contextual information is essential in covering a more comprehensive range of conflicts. For instance, to identify and resolve conflicts amongst various residents, an adaptive priority model is created while taking into account the contextual aspects of the residents (such as sickness, age, and impairment) \cite{chaki2021adaptive}. In addition, identifying different types of contextual information helps reveal unknown conflicts primarily ignored by current works. Therefore, we believe various contextual factors should be identified and considered in conflict detection and resolution.

\textbf{Severity level assessment:}  
Conflicts usually cause undesirable consequences and violate residents' expectations of enhanced life quality in smart homes \cite{chaki2021adaptive, chaki2021dynamic}. Therefore, it is necessary to assess the severity level of conflicts before resolving them. Some conflicts are severe and risky and need to be resolved immediately, while others are less serious and can be ignored. In worst cases, attackers can exploit conflicts hidden among IoT devices to achieve   malicious purposes. For example, two IoT apps aim at  managing indoor temperature. One app aims to turn on the heater when the room temperature is less than 20\degree C for comfortable living purposes. Another IoT app aims to open the window when the room temperature is more than 28\degree C for energy-saving purposes. Suppose an attacker has compromised the IoT network and has obtained   access to the heater. The conflict between the two IoT apps (i.e., The heater triggers the window to be opened through the temperature property.) provides the attacker with opportunities to break in easily. However, some conflicts are not much severe and can be ignored. For example, one IoT app may aim to turn  on the air-conditioning and set the temperature parameter to 22\degree C to cool the room when the temperature is more than 26\degree C. Another IoT app may aim to open the window to fresh air when the carbon dioxide is more than 0.5\%. Indeed, a conflict arises when turning on the air-conditioning while opening the window because opening the window would impact the air-conditioning's cooling effect. Sometimes, such a conflict may be unimportant and can be ignored. For example, suppose the resident's preferred temperature range is 22\degree C to 25\degree C. Although opening the window would allow outside hot air to come in and increase the room temperature to 24\degree C, the temperature value is still within the comfortable temperature range. Therefore, assessing the severity levels of conflicts is challenging, and extensive research is needed in this direction. 

\textbf{Preference extraction:} 
User preference is an important dimension of creating   smart living environments \cite{augusto2019user}. Extracting preferences towards the environment and  IoT device usage patterns plays a crucial role in detecting and resolving conflicts dynamically. Residents usually interact with IoT devices for various household chores, and these interactions can be recorded. We can extract users' appliance usage preferences by analyzing these interaction data. When people use a device/appliance frequently and periodically (a.k.a, consistently), that denotes their usage patterns \cite{choksi2022you}. These patterns generally reflect the residents' preferences \cite{wang2021attention}. However, occupants' usage patterns typically change and vary over time. For instance, a resident used to watch comedy movies every night at 8 pm but later started to watch drama every night at 8 pm. In this context, the resident's habits (i.e., TV-watching patterns) have changed. Therefore, there is a need to have a sophisticated preference extraction model that can dynamically capture residents' changes of preference. Preference extraction is the prerequisite of IoT service recommendation. In order to recommend/provide IoT services, a conflict may occur. Hence, conflict cannot be detected or resolved without appropriately extracting residents' preferences. Apart from the residents' preference for IoT device usage patterns, the preference towards environmental properties is also critical in detecting and resolving conflicts. The majority of the surveyed papers assume that environmental property is provided by residents or can be learned by various techniques \cite{liang2005thermal} \cite{demirezen2020development} \cite{serra2014smart}.

\textbf{Flexibility of preferences:}
The notion of conflict depends on the occupants' flexibility of preferences in   multi-resident smart homes. In practice, some residents are very accommodating (i.e., very flexible), whereas some are very strict (i.e., very inflexible) in terms of preferences. Suppose there are two residents in a home. On the one hand, one resident always watches a particular TV channel on Sundays between 7 pm and 8 pm. This depicts their strict nature of preference. On the other hand, another resident does not have a strict preference for TV channels. If they want to watch TV together at the same time and location, there is a high probability that a conflict may occur with the former resident if their preferred channel is not telecast. However, a conflict may not occur with the latter resident even though their preferred channel is not telecast as they do  not have any strict preference. Detecting conflicts beforehand is essential so that they can be resolved in advance. Therefore, it is necessary to identify the true nature of the residents' flexibility of preferences  to resolve the conflict. The conflict resolution approach for occupants with flexible preferences would differ from those with strict preferences. Indeed, the unpredictability of an occupant's daily life and their changing preferences over time make them subject to uncertainty. An apriori conflict detection approach is proposed in \cite{chaki2020fine}. However, they did not consider the rationale that combines both \emph{uncertainty} and the \emph{flexibility} of the residents' preferences. Therefore, a predictive conflict detection framework is required that reflects the residents' nature of flexibility and uncertainty regarding preferences.

\textbf{Conflict resolution:}
Enough attention to conflict resolution is not paid in the existing literature. As discussed earlier, the majority of work focuses on conflict detection. A conflict usually causes unwanted consequences and thus should be appropriately resolved. Conflict resolution aims to resolve disputes by selecting a course of action that will optimize resident satisfaction in terms of \emph{convenience}, \emph{effectiveness}, \emph{comfort}, \emph{safety}, and \emph{privacy} \cite{quijano2011user, lee2019situation}. Conflict resolution becomes an optimization problem in this regard. Therefore, designing the \emph{satisfaction} model and efficient optimization algorithm are two key tasks for resolving conflicts. Resolving conflicts is a challenging task. It needs to maximize the  \emph{satisfaction} score by considering a variety of aspects, including severity level as discussed above, priorities between different residents, system constraints,  and contexts (i.e., time and location information, as well as environment properties). There are a few preliminary works on resolving conflicts. To manage service conflicts, a locking method is suggested in \cite{hadj2017context}. When a service uses a shared resource, it merely locks the resource so that other services cannot use it.\looseness=-1

Appliance disputes are resolved in order of priority \cite{alfakeeh2016feature}. The resident assigns a priority value to each piece of equipment used in the services. Instead of completely blocking one service in favor of another, the negotiation mechanism aims to find a middle ground by enabling two incompatible services to operate side by side with slightly reduced efficiency. A dynamic conflict resolution strategy is devised by considering IoT devices' run-time behavior and disabling the device that violates the predefined policies \cite{celik2019iotguard, perumal2016rule, goynugur2017policy}. Similarly, the study   proposes to mitigate the attack chain by blocking one of the chained trigger-action rules \cite{hsu2019safechain}. However, these works present  a significant challenge in a complex IoT environment by simply disabling IoT services, which may result  in less satisfaction and incur unsafe and insecure states.  A negotiating agent approach is investigated to  help to implement an automatic interaction resolution for the feature interactions \cite{alfakeeh2022agent, alghamdi2015features, hsu2008smart}. An ontology-based approach is proposed to represent the environment and resolve users' conflicting environmental preferences \cite{camacho2014ontology}.

\section{Threats to Validity}
In this section, we discuss the potential threats to the validity of the research and the measures taken to mitigate these risks. We categorize three validity threats: (i) construct validity, (ii) internal validity, and (iii) external validity.

\subsection{Construct Validity}
A prevalent concern in systematic reviews is the potential exclusion of important scholarly works in the area of interest. This issue may arise for various reasons, such as a discrepancy between the keywords employed in academic work titles and abstracts and those used in contributions focusing on conflict detection in IoT-based smart homes. To mitigate threats to construct validity, the keywords ``conflict'' and ``smart home'' were established to form a dependable search string. Furthermore, these initial keywords were merged with terms like ``Internet of Things (IoT)'', ``IoT service'', ``conflict classification'', ``formal conflict model'', and ``conflict detection'' to encompass the widest range of scholarly works possible. Additionally, two reputable and widely recognized data sources were selected: ISI Web of Science and Scopus. Although limited to two databases, these sources optimize the pool of candidates, as both indexes the majority of scholarly works in prominent digital libraries, such as IEEE Xplore, ACM Digital Library, SpringerLink, and ScienceDirect. The snowballing technique was also employed to ensure the inclusion of articles that have a significant impact on the conflict detection study.

\subsection{Internal Validity}
Due to the numerous definitions for a single concept (i.e., conflict) and the lack of concise descriptions or suitable objectives and outcomes in certain scholarly works, there is a distinct risk of overlooking relevant studies or incorporating unrelated academic publications. Such constraints complicate the implementation of exclusion and inclusion criteria. To minimize errors stemming from subjective analysis, multiple meetings were conducted to address conflicts and establish appropriate resolutions. For instance, when uncertainty arose about the inclusion of a particular article, an initial meeting between the first two authors took place, followed by a subsequent meeting with advisors to reach a final decision. In this sub-section, we discuss the inclusion and exclusion criteria for selecting relevant studies.

The inclusion criteria defined are:
\begin{itemize}
    \item Any academic work that primarily addresses conflicts in an IoT-based environment that may cause IoT security and safety concerns. 
    \item Any academic work that was published in conferences or journals.
\end{itemize}

The exclusion criteria are:
\begin{itemize}
    \item The academic work that is not available on the web.
    \item The academic work which is not in English.
    \item The academic work that is primarily related to an area other than information systems, computer science, and
engineering.
    \item Academic works published in non-peer-reviewed conferences/journals.
    \item Academic works published before 2001.
\end{itemize}

\subsection{External Validity}
We consider that industrial, scientific, and academic communities in the smart homes research can benefit from this survey article.

\subsection{Study Selection}
We adopted PRISMA guidelines to select prior papers for this review \cite{moher2009preferred, page2021prisma}. Our search strategy, detailed in section 7.1, was built upon a comprehensive search string. This string was formulated by integrating key terms from relevant papers, their synonyms, alternative terms, and concepts related to conflict management in smart environments. We employed logical operators \textbf{AND} and \textbf{OR} for a more inclusive search. Preliminary searches were conducted to refine our search terms, ensuring the inclusion of known relevant papers. Our focus was on matching the search string with the titles, abstracts, and keywords of potential papers. The inclusion and exclusion criteria, elaborated in sections 7.2 and 7.3, further guided our paper selection. Searches across ACM Digital Library, IEEE Xplore, SpringerLink, and ScienceDirect yielded 151, 290, 46, and 37 papers, respectively. We meticulously reviewed the titles, abstracts, and keywords of these papers. In instances where the title and abstract were insufficient for a decision, the paper advanced to the next selection phase. After this initial screening, 290 papers remained. A thorough examination of the full text of these papers, applying our eligibility criteria, culminating in the selection of 52 papers specifically for conflict classification and detection.

\section{Conclusion}
Conflict is a natural phenomenon arising in an IoT-based smart home environment due to the customization performed by multiple residents on shared IoT devices  and  environments. Conflict usually results in undesirable situations and even serious security issues; thereby, it should be accurately detected and adequately resolved. Some models and approaches have been introduced to address the problem of conflict detection while lacking a systematic review of these approaches. This survey paper presents a systematic review of the latest research on conflict detection in an IoT-based smart home environment.   Overall, our survey contributes a conflict taxonomy to organize related works to reduce inconsistent understanding and confusion in the field. Additionally, it systematically analyses and compares recent conflict detection approaches to provide a big picture of how the challenges have been addressed and what progress has been made in the area of conflict detection. Therefore, our survey can provide researchers with a comprehensive understanding of the key aspects, main challenges, notable progress in this area and shed some light on future studies. This survey is expected to cover the majority of conflict types and can serve as a knowledge source to aid residents in personalizing conflict-free IoT-based services. It also facilitates security administrators to develop automatic conflict detection tools.

The detailed contributions of this survey are fourfold: (i) it presents a conflict taxonomy that defines and formalizes different types of conflicts based on the IoT rule model. The taxonomy has five types of conflicts, including Actuation Conflict, Preference Conflict, State Impact Conflict, and Direct and Indirect Environment Impact Conflict; (ii) it classifies conflicts based on the conflict taxonomy. This classification is refined under different contexts, including trigger-action rules, IoT apps, policies, IoT devices, IoT services, and home appliances. It provides a vision of current research on conflicts by identifying the detected conflicts and highlighting the uncovered conflicts  in each reviewed paper; (iii)  it comprehensively compares different conflict detection approaches and summarizes their contributions, and collects data sources. The conflict detection approaches are categorized into graph-based, object-oriented, formal rule modeling, model checking, and other approaches; (iv) it discusses the open issues and future directions in the field of IoT conflicts.   

In light of providing an intelligent home environment,  effectively detecting and resolving conflicts is a pressing problem in smart homes.  We identify five main research directions related to conflict detection and resolution. First, a complete conflict model is necessary for covering a more comprehensive range of conflicts that arise in different contexts. An evaluation schema is needed to assess the risk level of the detected conflicts before resolving them. Third, a robust preference extraction model that can dynamically capture residents' preferences changes. Fourth, a preference flexibility model is required that reflects the residents' nature of flexibility and uncertainty regarding preferences. Fifth, an effective conflict resolution approach is required to appropriately mitigate different types of conflicts to maximize residents' overall satisfaction.

A shortcoming of this survey is that the conflict taxonomy covers the majority of conflicts but does not cover all types of conflict. This is because each conflict model captures the key and common characteristics of similar conflicts arising in different contexts.  As a result, some special cases of conflicts may be overlooked. As aforementioned discussion, the less coverage issue of our conflict model can be simply solved by scaling the model with more features of IoT devices at the cost of increasing the difficulty in comprehension for readers.

\bibliographystyle{ACM-Reference-Format}
\bibliography{ConfDet}

\appendix

\end{document}